\documentclass[english,10pt,aps,prd,a4paper,preprintnumbers,floatfix,nofootinbib,showpacs,superscriptaddress, notitlepage]{revtex4-1}

\pdfoutput=1
\usepackage{latexsym}
\usepackage{amsthm}
\usepackage{amsmath}
\usepackage[dvips]{graphicx}
\usepackage{wrapfig}
\usepackage{amssymb}
\usepackage{color}
\usepackage{cancel}
\usepackage{multirow}
\usepackage{pst-3d}                  
\usepackage{pst-grad}                
\usepackage{pst-node}                
\usepackage{pst-slpe}                
\usepackage{pst-plot}                %
\usepackage{pst-coil}                %
\usepackage{pst-math}                %
\usepackage{pstricks-add}            %
\usepackage{rotating}
\usepackage{geometry}
\usepackage{hhline}
\usepackage{amsmath}
\usepackage{float}
\usepackage{ulem}                    
\usepackage{fancybox}
\usepackage[utf8]{inputenc}
\usepackage[colorlinks=true,citecolor=darkred,urlcolor=darkred, pdfborder={0 0 0}]{hyperref}
\definecolor{darkred}{rgb}{0.6,0,0}
\usepackage{braket}

\allowdisplaybreaks
\allowdisplaybreaks[1]
\allowdisplaybreaks[2]

\newcommand{\f}{\!\!\!\!}

\definecolor{greenLinks}{rgb}{0, 0.6, 0} 
\definecolor{blueLinks}{rgb}{0, 0, 0.6}
\definecolor{redLinks}{rgb}{0.6, 0, 0}
\definecolor{tempText}{rgb}{0.55, 0.10,0.67}
\definecolor{eprintLinks}{rgb}{0.4, 0.4, 0.4}
\definecolor{journalLinks}{rgb}{0.6, 0, 0}
\newcommand {\ignore}[1]{}

\def\cpv{CP violation }

\newcommand{\sm}{{Standard Model }}

\def\lsim{\mathrel{\rlap{\lower4pt\hbox{\hskip1pt$\sim$}}
    \raise1pt\hbox{$<$}}}
\def\gsim{\mathrel{\rlap{\lower4pt\hbox{\hskip1pt$\sim$}}
    \raise1pt\hbox{$>$}}}

\def\U1s{$\mathrm{U_{1}^{(a)}\otimes U_{1}^{(b)}\otimes U_{1}^{(c)}\otimes U_{1}^{(d)}\otimes U_{1}^{(e)}}$ }
\def\3211{$\mathrm{SU(3) \otimes SU(2)_L \otimes U(1)_R \otimes U(1)_{B-L}}$ }
\def\321{$\mathrm{SU(3) \otimes SU(2) \otimes U(1)}$ }
\def\422{$\mathrm{SU(4) \otimes SU(2) \otimes SU(2)_R}$ }

\newcommand{\AddrAHEP}{%
  AHEP Group, Institut de F\'{i}sica Corpuscular --
  C.S.I.C./Universitat de Val\`{e}ncia, Parc Cient\'ific de Paterna.\\
 C/ Catedr\'atico Jos\'e Beltr\'an, 2 E-46980 Paterna (Valencia) - SPAIN}

\begin{document}

\begin{flushright}
\ \hfill\mbox{\small USTC-ICTS-18-21}\\[4mm]
\begin{minipage}{0.2\linewidth}
\normalsize
\end{minipage}
\end{flushright}
\bibliographystyle{unsrt}   

\title{CP Symmetries as Guiding Posts: Revamping Tri-Bi-Maximal Mixing-I}
\author{Peng Chen}\email{pche@mail.ustc.edu.cn}
\affiliation{College of Information Science and Engineering,Ocean University of China, Qingdao 266100, China}
\author{Salvador Centelles Chuli\'{a}}\email{salcen@ific.uv.es}
\affiliation{\AddrAHEP}
\author{Gui-Jun Ding}\email{dinggj@ustc.edu.cn}
\affiliation{Interdisciplinary Center for Theoretical Study and Department of Modern Physics, \\
University of Science and Technology of China, Hefei, Anhui 230026, China}
\author{Rahul Srivastava}\email{rahulsri@ific.uv.es}
\affiliation{\AddrAHEP}
\author{Jos\'{e} W. F. Valle}\email{valle@ific.uv.es}
\affiliation{\AddrAHEP}

\begin{abstract}
  \vspace{1cm}

We analyze the possible generalized CP symmetries admitted by the Tri-Bi-Maximal (TBM) neutrino mixing. Taking advantage of these symmetries we construct in a systematic way other variants of the standard TBM ansatz. Depending on the type and number of generalized CP symmetries imposed, we get new mixing matrices, all of which related to the original TBM matrix. One of such ``revamped'' TBM variants is the recently discussed 
mixing matrix of arXiv:1806.03367.  We also briefly discuss the phenomenological implications following from these mixing patterns.

\end{abstract}

\maketitle


\section{Introduction}
\label{sec:introduction}

The historic discovery of neutrino oscillations~\cite{Kajita:2016cak,McDonald:2016ixn} marked the beginning of a new era in particle physics~\cite{Valle:2015pba} in which it becomes manifest that the \sm needs amendment.
Many basic drawbacks in cosmology associated with the origin of matter and the evolution of the universe also point in the same direction.
According to the Big-Bang, the early Universe would have created equal amounts of matter and antimatter.
Yet there is an overwhelming dominance of matter in the universe. This indicates that matter must behave rather differently from antimatter.
Indeed such observed asymmetry may be the result, among other things, of the existence of CP violation in nature. Within the perturbative \sm picture \cpv exists only in the quark sector. However, CP violation present in the quark sector does not seem enough to account for the observed matter to anti-matter asymmetry within the Standard Model~\cite{Dine:2003ax}.
The structure of the lepton sector and the properties of neutrinos come forward as possible key ingredients in the resolution of this dilemma, through the mechanism of leptogenesis~\cite{Fukugita:1986hr}.
Indeed, recent neutrino oscillation global studies provide a first hint for CP violation in the lepton sector~\cite{deSalas:2017kay}. Upcoming neutrino oscillation experiments aim to improve our understanding of neutrinos through the precise measurement of leptonic CP violation~\cite{Acciarri:2015uup,Srivastava:2018ser,Nath:2018fvw}.
If neutrinos are, as expected in many theories, self-conjugate fermions, then there are also Majorana phases characterizing CP violation in the lepton sector~\cite{Schechter:1980gr}.
While these phases do not affect oscillations, they are crucial in the description of neutrinoless double-beta decay~\cite{Schechter:1980gk}. The current experimental sensitivities are given in~\cite{KamLAND-Zen:2016pfg,GERDA:2018,MAJORANA:2008,CUORE:2018,EXO-2018,Arnold:2016bed}. 

It is therefore of fundamental importance to make theoretical predictions for neutrino mixing parameters as well as CP violating phases. The most reasonable and popular approach is to appeal to symmetry considerations~\cite{Ishimori:2010au}.
Rather than considering specific theories on a one-by-one basis, here we adopt a more model-independent theory framework based on the imposition of residual CP symmetries, irrespective of how the relevant mass matrices arise from first principles~\cite{Chen:2014wxa,Chen:2015siy,Chen:2016ica,Chen:2018lsv}. As a starting point we take a complexified version of the standard Tri-Bi-Maximal (TBM) neutrino mixing ansatz~\cite{harrison:2002er} as our benchmark neutrino mixing pattern. 
There are three independent generalized CP symmetries admitted by the TBM ansatz, if all of them are imposed we recover the starting point. However if we partially impose the generalized CP symmetry we can construct other non-trivial variants of the standard TBM ansatz in a systematic manner. 
Depending on the type and number of generalized CP symmetries imposed, we get several different mixing matrices. Such ``revamped'' TBM variants have in general non-zero $\theta_{13}$ as well as CP violation, as currently indicated by the experimental data. A simple example of this procedure has already been given in~\cite{Chen:2018eou}.

This paper is structured as follows. In Section~\ref{sec:preliminaries} we give a general preliminary discussion of the method,
while in  Section~\ref{sec:cp-symm-trib} we describe the CP symmetries of tribimaximal mixing. We then move on to describe the neutrino 
mass matrices conserving two and one CP symmetries, in Sections~\ref{sec:two-neut} and \ref{sec:one-cp}, respectively. 
We show that these mass matrices lead to realistic mixing matrices with non-zero $\theta_{13}$ which are closely related to the TBM matrix and share some of its properties. We also discuss the phenomenological predictions from these matrices.
Finally, we give a brief sum-up discussion at the end.


\section{Preliminaries}
\label{sec:preliminaries}


In this section we begin with a general discussion of the generalized CP transformations, highlighting key concepts as well as setting 
up our notation and conventions. Following Refs.~\cite{Chen:2014wxa,Chen:2015siy,Chen:2016ica} we start by defining the generalized 
remnant CP transformations for each fermionic field as follows:
\begin{eqnarray}
 \psi \stackrel{CP}{\longmapsto} i X_\psi \gamma ^0 \mathcal{C} \bar{\psi} ^T, \hspace{2mm} \psi \in \{\nu_L, \nu_R, l_L, l_R \}\,.
\end{eqnarray}
Such generalized CP transformations acting on the chiral fermions will
be a symmetry of the mass term in the Lagrangian provided they satisfy
the following conditions~\footnote{Even though the X-matrix is symmetric,
  we prefer to use $X^\dagger$ instead of $X^*$ when dealing with Dirac fields.},
\begin{eqnarray}
X_\psi^Tm_\psi X_\psi & = & m_\psi^*,\quad \textnormal{ for Majorana fields}\,, \label{eq:cpmaj}   \\
X_\psi^\dagger M^2_\psi X_\psi & = & M_\psi^{2 *},\quad \textnormal{ for Dirac fields, where } M^2_\psi \equiv m_\psi^\dagger \, m_\psi \,, \label{eq:cpdir}
\end{eqnarray}
where $m_{\psi}$ is written in a basis with left-handed (right-handed) fields on the right-hand (left-hand) side. 
Note that the mass matrices $m_{\psi}$ and $M^2_{\psi}$ can be diagonalized by a unitary transformation $U_{\psi}$,

\begin{eqnarray}
\label{eq:diag_maj}&&U^T_{\psi} m_{\psi} \,U_{\psi}=\text{diag}(m_1,m_2,m_3),\quad \textnormal{ for Majorana fields}\,,\\
\label{eq:diag_dir}&&U^\dagger_{\psi} M^2_{\psi} U_{\psi}=\text{diag}(m^2_1,m^2_2,m_3^2),\quad \textnormal{for Dirac fields}\,,
\end{eqnarray}
with $m_1\neq m_2\neq m_3$.  From Eqs.~\eqref{eq:cpmaj}-\eqref{eq:diag_dir}, after straightforward algebra, we find that the unitary transformation $U_{\psi}$ is subject to the following constraint on the imposed CP symmetry $X_{\psi}$,
\begin{equation}
\label{eq:XCons}
U^\dagger_{\psi} X_{\psi} U^*_{\psi}\equiv P=\left\{\begin{array}{cc}
\text{diag}(\pm 1,\pm 1,\pm 1),& \textnormal{ for Majorana fields},\\[0.1in]
\text{diag}(e^{i\delta_e}, e^{i\delta_\mu}, e^{i\delta_\tau}),& \textnormal{for Dirac fields}\,,
\end{array}
\right.
\end{equation}
where $\delta_{e}$, $\delta_{\mu}$ and $\delta_{\tau}$ are arbitrary real parameters~\footnote{If neutrinos are Majorana particles and the
  lightest one is massless (this possibility is still allowed by current experimental data) one ``$\pm$'' entry would be a complex
  phase.}. Because $X_{\psi}$ is a symmetric matrix, one can use Takagi decomposition (note that this decomposition is not unique) to
express $X_{\psi}$ as
\begin{eqnarray}
\label{eq:Td}X_{\psi}=\Sigma \cdot \Sigma^T\,.
\end{eqnarray}
Inserting Eq.~\eqref{eq:Td} into Eq.~\eqref{eq:XCons}, we find that the combination $P^{-\frac{1}{2}} U^{\dagger}_{\psi}\Sigma$ is a real
orthogonal matrix, i.e.,
\begin{eqnarray}
\label{eq:ugen}P^{-\frac{1}{2}}U^{\dagger}_{\psi}\Sigma \, = \, O_{3\times3}\,,
\end{eqnarray}
which implies
\begin{equation}
\label{eq:ugenn}
U_{\psi}=\Sigma O^{T}_{3\times3} P^{-\frac{1}{2}}\,,
\end{equation}
where $O_{3\times3}$ is a generic $3\times 3$ real orthogonal matrix. Note that, with appropriate $P$, this holds equally well if neutrinos are Majorana
or Dirac--type~\cite{Chen:2018lsv}. In the latter case the Majorana phases are unphysical and neutrinoless double beta decay is forbidden.

Note that Eq.~\eqref{eq:ugenn} may be regarded as a prediction for the lepton mixing matrix $U_{\psi}$ associated to the given residual
 CP symmetry encoded in $X_{\psi}$ or $\Sigma$~\cite{Chen:2014wxa,Chen:2015siy,Chen:2016ica, Chen:2018lsv}~\footnote{ There is a more intuitive way of deriving Eq.~\eqref{eq:ugenn}, which will be given later on.}.
At this point we would like to remark that the generalized CP symmetries do not impose any constraint on the fermion masses. These can always be chosen to match the required experimental values. The predictive power of generalized CP symmetries lies in the mixing matrix elements and their phases.
In this paper we will use the predictive power of Eq.~\eqref{eq:ugenn} in a different way by explicitly building the mass matrices that satisfy a certain CP symmetry. 
This offers a more intuitive procedure useful for model building.

\section{CP symmetries of tribimaximal mixing}
\label{sec:cp-symm-trib}

For a long time the TBM mixing pattern has been a popular ansatz for the possible structure of the lepton mixing matrix.
In what follows we will exploit the predictive power of generalized CP symmetries in a different way. Rather than predicting lepton mixing as in Eq.~\eqref{eq:ugenn}, we will assume that the TBM provides a good starting point and derive the possible deviations from that benchmark ansatz in a way consistent with the assumed generalized CP symmetries.

The standard TBM pattern leads to three predictions, given as
\begin{equation}
\theta_{12}=\arcsin\frac{1}{\sqrt{3}}\,,\qquad\qquad
\theta_{23}=\frac{\pi}{4}\,,\qquad\qquad
\theta_{13}=0\,.
\label{eq:tbm-angles}
\end{equation}
Owing to the fact that $\theta_{13} = 0$, the Dirac CP phase $\delta_{CP}$ becomes unphysical. 
In its simplest form the TBM mixing was assumed to be completely CP conserving and thus both Majorana phases were also set to zero.
We will refer to this CP conserving TBM matrix as the ``real TBM'' mixing matrix, given by
 \begin{eqnarray}
 {U_{rTBM}} & = &  \left[
\begin{array}{ccc}
\sqrt{\frac{2}{3}}               & \frac{1}{\sqrt{3}}           & 0   \\
-\frac{1}{\sqrt{6}}              & \frac{1}{\sqrt{3}}           & \frac{1}{\sqrt{2}} \\
\frac{1}{\sqrt{6}}               &  -\frac{1}{\sqrt{3}}         & \frac{1}{\sqrt{2}} \\
\end{array}
\right]
\label{eq:tbm}
\end{eqnarray}

However, if neutrinos are Majorana in nature we can assume Majorana phases to be nonzero. We call the resulting mixing matrix for such case as ``complex TBM'' matrix (cTBM)~\cite{Chen:2018eou} and its form is given by
\begin{eqnarray}
 U_{cTBM} & = &  \left[
\begin{array}{ccc}
\sqrt{\frac{2}{3}}                       & \frac{ e^{-i \rho } }{\sqrt{3}}     & 0                                \\
-\frac{e^{i \rho }}{\sqrt{6}}            & \frac{1}{\sqrt{3}}                  & \frac{e^{-i\sigma}}{\sqrt{2}}    \\
\frac{e^{i (\rho + \sigma)}}{\sqrt{6}}   &  -\frac{e^{i\sigma}}{\sqrt{3}}      & \frac{1}{\sqrt{2}}               \\
\end{array}
\right]
\label{eq:ctbm}
\end{eqnarray}
The cTBM mixing matrix in Eq.~\eqref{eq:ctbm} predicts the same mixing angles as real TBM in Eq.~\eqref{eq:tbm-angles}, but now the 
Majorana phases are nonzero. In the symmetrical parametrization~\cite{Schechter:1980gr,Rodejohann:2011vc} they are given by
\begin{eqnarray}
 \phi_{12} & = & \rho \, \qquad \phi_{23} \, = \, \sigma
\end{eqnarray}
Given the recent oscillation measurements~\cite{An:2016ses,Pac:2018scx,Abe:2014bwa}, neither the real nor the complex variant
of the TBM ansatz is a viable lepton mixing pattern.

In this work we show that starting from the cTBM matrix, and using the methodology of generalized CP symmetries, one can systematically construct and analyze realistic neutrino mixing matrices with non-zero reactor angle.
These mixing matrices share many other properties of the simplest TBM mixing matrix.

In order to illustrate our methodology we assume neutrinos to be Majorana-type, and start with the complex TBM matrix of Eq.~\eqref{eq:ctbm}. 
The real TBM matrix can always be obtained from it by simply taking the limit $\rho, \sigma \to 0$.
In what follows we will take this limit at various points of our discussion.
Moreover, for sake of simplicity, throughout this paper we work in the charged lepton diagonal basis.

Let us start our discussion by inverting Eq.~\eqref{eq:XCons}, so as to obtain the four CP symmetry matrices $X_i$ associated with the cTBM ansatz. These are given by~\cite{Chen:2014wxa,Chen:2015nha}
\begin{eqnarray}
X_i = U_{cTBM}\,\hat{d}_i\,U_{cTBM}^T\,,
\qquad\text{where}\qquad\quad
\begin{aligned}
&\hat{d}_1=\text{diag}\left(1,-1,-1\right),\qquad \hat{d}_2=\text{diag}\left(-1,1,-1\right),\\
&\hat{d}_3=\text{diag}\left(-1,-1,1\right),\qquad \hat{d}_4=\text{diag}\left(1,1,1\right)\,.
\end{aligned}
\label{eq:defX}
\end{eqnarray}
In matrix form these four CP symmetries can be written as
\begin{eqnarray}
\nonumber X_1 & = & \frac{1}{6}
\left(\begin{array}{ccc}
 4-2 e^{-2 i \rho } & -2 e^{-i \rho }-2 e^{i \rho } & 2 e^{i (\rho + \sigma) }+2 e^{-i ( \rho - \sigma) } \\
 -2 e^{-i \rho }-2 e^{i \rho } & -2+e^{2 i \rho }-3 e^{-2 i \sigma } & -3 e^{-i \sigma }-e^{i (2\rho + \sigma) }+2 e^{i \sigma } \\
 2 e^{i (\rho + \sigma) }+2 e^{-i ( \rho - \sigma )} & -3 e^{-i \sigma }-e^{i (2\rho + \sigma)  }+2 e^{i \sigma} & -3+e^{2 i (\rho + \sigma)} -2 e^{2 i \sigma } \\
\end{array}\right)\,,\\
\nonumber X_2 & = & \frac{1}{6}
\left(\begin{array}{ccc}
 -4+2 e^{-2 i \rho } & 2 e^{-i \rho }+2 e^{i \rho } & -2 e^{i (\rho + \sigma)  }-2 e^{-i ( \rho - \sigma) } \\
 2 e^{-i \rho }+2 e^{i \rho } & 2-e^{2 i \rho }-3 e^{-2 i \sigma } & -3 e^{-i \sigma }+e^{i (2\rho + \sigma) }-2 e^{i \sigma} \\
 -2 e^{i (\rho + \sigma)  }-2 e^{-i ( \rho - \sigma ) } & -3 e^{-i \sigma }+e^{i (2\rho + \sigma) }-2 e^{i \sigma } &
 -3  - e^{2 i (\rho + \sigma) } +2 e^{2 i \sigma} \\
\end{array}\right)\,,\\
\nonumber X_3 & = & \frac{1}{6}
\left(\begin{array}{ccc}
 -4-2 e^{-2 i \rho } & -2 e^{-i \rho }+2 e^{i \rho } & -2 e^{i (\rho + \sigma) }+2 e^{-i ( \rho - \sigma) } \\
 -2 e^{-i \rho }+2 e^{i \rho } & -2-e^{2 i \rho }+3 e^{-2 i \sigma } & 3 e^{-i \sigma }+e^{i (2\rho + \sigma) }+2 e^{i \sigma } \\
 -2 e^{i (\rho + \sigma)  }+2 e^{-i ( \rho - \sigma )} & 3 e^{-i \sigma }+e^{i (2\rho + \sigma)  }+2 e^{i \sigma } & 3-e^{2 i (\rho + \sigma)  }-2 e^{2 i \sigma } \\
\end{array}\right)\,,\\
\label{eq:X0} X_4 & = & \frac{1}{6}
\left(\begin{array}{ccc}
 4+2 e^{-2 i \rho } & 2 e^{-i \rho }-2 e^{i \rho } & 2 e^{i (\rho + \sigma) }-2 e^{-i ( \rho - \sigma) } \\
 2 e^{-i \rho }-2 e^{i \rho } & 2+e^{2 i \rho }+3 e^{-2 i \sigma } & 3 e^{-i \sigma }-e^{i (2\rho + \sigma) }-2 e^{i \sigma } \\
 2 e^{i (\rho + \sigma) }-2 e^{-i ( \rho - \sigma )} & 3 e^{-i \sigma }-e^{i (2\rho + \sigma) }-2 e^{i \sigma } & 3+e^{2 i (\rho + \sigma)  }+2 e^{2 i \sigma} \\
\end{array}\right)\,.
\label{eq:ccp-sym}
\end{eqnarray}
The CP symmetries corresponding to the real TBM matrix of Eq.~\eqref{eq:tbm} are obtained simply by taking the limit of $\rho, \sigma \to 0$ in Eq.~\eqref{eq:ccp-sym}. 
These CP symmetries in matrix form are given by
\begin{eqnarray}
X_1=\frac{1}{3}\left(
\begin{array}{ccc}
\!\!  1 \!&\!  -2  \!&\!  2 \!\!\\
\!\! -2 \!&\!  -2  \!&\! -1 \!\!\\
\!\!  2 \!&\! -1  \!&\!  -2 \!\!
\end{array}\right)\,,\quad
X_2=\frac{1}{3}\left(
\begin{array}{ccc}
\!\! -1 \!&\!  2 \!&\! -2 \!\!\\
\!\!  2 \!&\! -1 \!&\! -2 \!\!\\
\!\! -2 \!&\! -2 \!&\! -1 \!\!
\end{array}\right)\,,\quad
X_3=\left(\begin{array}{ccc}
\!-1 \!&\! 0 \!&\! 0 \!\\
\! 0 \!&\! 0 \!&\! 1 \!\\
\! 0 \!&\! 1 \!&\! 0 \!
\end{array}\right)\,,\quad
X_4=\left(\begin{array}{ccc}
\! 1 \!&\! 0 \!&\! 0 \!\\
\! 0 \!&\! 1 \!&\! 0 \!\\
\! 0 \!&\! 0 \!&\! 1 \!
\end{array}\right)\,.
\label{eq:rcp-sym}
\end{eqnarray}
Notice that the $X_3$ in Eq.~\eqref{eq:rcp-sym} is nothing but the famous $\mu - \tau$ reflection 
symmetry~\cite{harrison:2002er,Babu:2002dz, Grimus:2003yn}, characteristic of the real TBM matrix, while $X_4$ is simply the identity CP symmetry. Moreover, all the four above CP transformations in Eq.~\eqref{eq:rcp-sym} can be be reproduced from the breaking of $S_4$ flavor symmetry and generalized CP symmetry~\cite{Feruglio:2012cw,Ding:2013hpa,Chen:2016ptr}.

Analogously there are four flavor symmetry transformations $G_i$ associated with the cTBM ansatz~\cite{Chen:2014wxa,Chen:2015nha}
\begin{eqnarray}
\label{eq:CP_Fav}
\begin{aligned}
&G_1 = X_2X^{\ast}_3=X_3X^{\ast}_2=X_4X^{\ast}_1=X_1X^{\ast}_4\,,\qquad\qquad G_2 = X_1X^{\ast}_3=X_3X^{\ast}_1=X_4X^{\ast}_2=X_2X^{\ast}_4\,,\\
&G_3 = X_1X^{\ast}_2=X_2X^{\ast}_1=X_4X^{\ast}_3=X_3X^{\ast}_4\,,\qquad\qquad G_4 = X_1X^{\ast}_1=X_2X^{\ast}_2=X_3X^{\ast}_3=X_4X^{\ast}_4\,.
\end{aligned}
\end{eqnarray}

Notice that out of the four CP and flavor symmetries only three are really independent~\cite{Chen:2014wxa, Chen:2015nha}.
If any three of the four CP symmetries in Eq.~\eqref{eq:X0} are imposed simultaneously then one can uniquely reconstruct the neutrino mixing matrix
which will be nothing but cTBM of Eq.~\eqref{eq:ctbm} which $\theta_{13} = 0$ and hence fails to provide a viable description of lepton mixing~\cite{An:2016ses, Pac:2018scx, Abe:2014bwa,deSalas:2017kay}.
However, as we will discuss in rest of this paper, imposing only two or one CP symmetry can lead to realistic mixing patterns with non-zero reactor angle and CP violation in oscillations. The latter results from the incomplete CP symmetry of the transformation matrices defining the corresponding theories. \\


\section{Neutrino mass matrix conserving two CP symmetries}
\label{sec:two-neut}


Our goal in what follows is to derive various variants of the TBM mixing pattern that can be 
obtained when the neutrino mass matrix respects only a partial set of the above CP symmetries.
We start our discussion by looking at the case when the neutrino mass matrix respects only two CP symmetries of the four given in Eq.~\eqref{eq:ccp-sym}.
Although no longer strictly viable, given the reactor measuremts of $\theta_{13}$ as well as the first hints 
for nonzero $\delta_{CP}$ from oscillation experiments, the TBM matrix still provides a good zero-th order 
approximations that captures the main features of lepton mixing. Hence it provides a valid benchmark that we
can perturb slightly, in a controlled way, subject to CP symmetry requirements.

In this section we analyze neutrino mass matrices which preserve, at leading order, all four CP symmetries of Eq.~\eqref{eq:ccp-sym}.
To obtain realistic mass and mixing matrices, in addition to this leading order mass term we add perturbation 
terms which will preserve only two CP symmetries of the complex TBM matrix.
We now consider the various cases when the perturbation term preserves only two of the CP symmetries 
of the Majorana neutrino mass matrix. The addition of perturbation terms preserving fewer symmetries 
in turn implies that the leptonic mixing matrix is no longer the cTBM matrix but a closely related sibling.
For definiteness we work in the basis in which the charged lepton mass matrix is diagonal, 
so that the leptonic mixing matrix is described solely by the neutrino mixing matrix.
As already mentioned, realistic variants of the real TBM matrix of Eq.~\eqref{eq:tbm} can be obtained simply by taking the limit $\rho, \sigma \to 0$.

Having said that we start, at the leading order, by requiring that the neutrino mass matrix $M_\nu^{(0)}$ satisfies
\begin{equation}
X_i^T M_\nu^{(0)} X_i = M_\nu^{(0)\,\ast}\,,
\end{equation}
where $X_i$; $i=1,2,3,4$ are the four CP symmetries of Eq.~\eqref{eq:ccp-sym}. Thus for $M_\nu^{(0)}$ we have
\begin{equation}
\hat{d}_i \, U_{cTBM}^T M_\nu^{(0)} U_{cTBM}\, \hat{d}_i = (U_{cTBM}^T M_\nu^{(0)} U_{cTBM})^\ast\,.
\end{equation}
This in turn implies that $U_{cTBM}^T M_\nu^{(0)} U_{cTBM}$ is a real diagonal matrix. Thus we can write it as
\begin{equation}
U_{cTBM}^T M_\nu^{(0)} U_{cTBM}=\text{diag}(m_1,m_2,m_3)\,,\qquad\qquad
M_\nu^{(0)}=U_{cTBM}^\ast\text{diag}(m_1,m_2,m_3)U_{cTBM}^\dagger\,.
\label{eq:m0}
\end{equation}

Now we add to this leading term a perturbation term $\delta M_\nu$ which only preserves two of the four CP symmetries.
There are six different possible pairs of CP symmetries that can be preserved by $\delta M_\nu$, namely,
\begin{eqnarray}
(X_1,X_4)\,,\qquad(X_2,X_3)\,,\qquad(X_2,X_4)\,,\qquad(X_1,X_3)\,,\qquad(X_3,X_4)\,,\qquad(X_1,X_2)\,.
\label{eq:two-sym-combo}
\end{eqnarray}
Since, as seen in Eq.~\eqref{eq:CP_Fav}, the CP symmetries are also related with the flavor symmetries, it follows that
 preserving two CP symmetries also implies that certain flavor symmetries of cTBM are preserved even by the perturbation term.
The flavor and CP symmetries that are preserved can be grouped as
\begin{eqnarray}
(X_1\,,\,X_4)\quad\text{or}\quad(X_2\,,\,X_3)\qquad&\Rightarrow&\qquad \text{$G_1$ flavor symmetry preserved;} \nonumber \\
(X_2\,,\,X_4)\quad\text{or}\quad(X_1\,,\,X_3)\qquad&\Rightarrow&\qquad \text{$G_2$ flavor symmetry preserved;} \nonumber \\
(X_3\,,\,X_4)\quad\text{or}\quad(X_1\,,\,X_2)\qquad&\Rightarrow&\qquad \text{$G_3$ flavor symmetry preserved.}
\label{eq:cp-flav-combo}
\end{eqnarray}
In the following subsections we discuss the first two cases in detail~\footnote{One can check that models based on the $G_3$ symmetry are not experimentally viable since the reactor angle $\theta_{13}$  remains zero.}. We will see how the incomplete 
imposition of the CP symmetry in the corresponding transformation matrices characterizing each theory 
can result in realistic mixing patterns with non-zero reactor angle as well as CP violation in oscillations.

\subsection{$G_1$ flavor and $X_1, X_4$ CP symmetries preserved}
\label{sec:cp14-g1}

We first consider the case when the perturbation term preserves the $X_1$ and $X_4$ CP symmetries. 
This also implies that the $G_1$ flavour symmetry is preserved for this case, so that the perturbation term satisfies
\begin{eqnarray}
X_i^T \delta M_\nu X_i=\delta M_\nu^{\ast}\,,\qquad
\hat{d}_i \, U_{cTBM}^T\delta M_\nu  U_{cTBM}\, \hat{d}_i = (U_{cTBM}^T\delta M_\nu  U_{cTBM})^\ast\,.
\end{eqnarray}
with $i=1$ and $4$. Thus, $U_{cTBM}^T\delta M_\nu  U_{cTBM}$ must be the form 
\begin{equation}
U_{cTBM}^T \delta M_\nu U_{cTBM} =
\left(\begin{array}{ccc}
 \delta m_1^\prime & 0 & 0 \\
 0 & \delta m_2^\prime & \delta m \\
 0 & \delta m & \delta m_3^\prime
\end{array}\right)\,.
\end{equation}
Since $\delta m_1^\prime$, $\delta m_2^\prime$ and $\delta m_3^\prime$ can always be absorbed by $m_1$, $m_2$ and $m_3$ of $M_\nu^{(0)}$ in Eq.~\eqref{eq:m0}, we can take $\delta m_1^\prime = \delta m_2^\prime = \delta m_3^\prime = 0$ without loss of generality.
Thus $\delta M_\nu$ simplifies to 
\begin{equation}
\delta M_\nu =
U_{cTBM}^T\left(\begin{array}{ccc}
 0 & 0 & 0 \\
 0 & 0 & \delta m \\
 0 & \delta m & 0
\end{array}\right)
\,.
\end{equation}
Thus the full mass matrix ($M_\nu = M_\nu^{(0)} + \delta M_\nu$) satisfies the following relation  
\begin{equation}
\label{eq:hatm14}
U_{cTBM}^T (M_\nu^{(0)}+\delta M_\nu)U_{cTBM} =
\left(\begin{array}{ccc}
 m_1 & 0 & 0 \\
 0 & m_2 & \delta m \\
 0 & \delta m & m_3
\end{array}\right)\,.
\end{equation}
Owing to the off-diagonal perturbation term $\delta m$ the full mass matrix is no longer diagonalized by the $U_{cTBM}$ matrix alone.  
However, Eq.~\eqref{eq:hatm14} can be easily diagonalized by an orthogonal matrix $O_{23}$ given by
\begin{equation}
O_{23}=
\left(\begin{array}{ccc}
1 & 0 & 0 \\
0 & \cos\theta & \sin\theta \\
0 & -\sin\theta & \cos\theta
\end{array}\right)\,,
\qquad\text{with}\qquad
\begin{aligned}
&\tan2\theta=\frac{2\delta m}{m_3-m_2}\,,\\
&\cos2\theta=\frac{m_3-m_2}{\sqrt{(m_3-m_2)^2+4\delta m^2}}\,.
\end{aligned}
\end{equation}
Thus we have
\begin{equation}
\label{eq:full-m-14}
O_{23}^T U_{cTBM}^T (M_\nu^{(0)}+\delta M_\nu) U_{cTBM} O_{23} = \text{diag} (m_1', m_2', m_3') \,.
\end{equation}
The mass eigenvalues are given by
\begin{eqnarray}
m_1^\prime=m_1, \quad
m_2^\prime=\frac{1}{2}\big(m_2+m_3-\sqrt{(m_3-m_2)^2+4\delta m^2}\big),\quad
m_3^\prime=\frac{1}{2}\big(m_2+m_3+\sqrt{(m_3-m_2)^2+4\delta m^2}\big)\,.
\end{eqnarray}
Since we are working in the diagonal charged lepton mass basis, the full leptonic mixing matrix $U_{lep}$ is
simply given by the neutrino mixing matrix.
Thus we have
\begin{equation}
\label{eq:ux14}U_{lep} = U_{cTBM}\,O_{23}\,Q_{\nu}\, \, = \,
\left(\begin{array}{ccc}
\sqrt{\frac{2}{3}}
& \frac{e^{-i \rho} \cos \theta}{\sqrt{3}}
&  \frac{e^{-i \rho} \sin \theta}{\sqrt{3}}
       \\
-\frac{e^{i \rho}}{\sqrt{6}}
& \frac{\cos \theta}{\sqrt{3}} - \frac{e^{-i \sigma} \sin \theta}{\sqrt{2}}
&  \frac{\sin \theta}{\sqrt{3}} + \frac{e^{-i \sigma} \cos \theta}{\sqrt{2}}
       \\
\frac{e^{i (\rho + \sigma)}}{\sqrt{6}}
& - \frac{\sin \theta}{\sqrt{2}} - \frac{e^{i \sigma} \cos \theta}{\sqrt{3}}
&  \frac{\cos \theta}{\sqrt{2}} - \frac{e^{i \sigma} \sin \theta}{\sqrt{3}}
\end{array}\right)\,Q_{\nu} \,.
\end{equation} 
where $Q_{\nu} = \text{diag} (e^{i k_1 \pi /2}, e^{i k_2 \pi /2} , e^{i k_3 \pi /2} )$ is a diagonal matrix of phases.

From Eq.~\eqref{eq:ux14} one can easily extract the parameters characterizing the lepton mixing matrix $U_{lep}$ which, in symmetric parametrization, are given by
\begin{eqnarray}
\nonumber\sin^2\theta_{13} & = & \frac{\sin^2\theta}{3}\,,\qquad\qquad \sin^2\theta_{12}=\frac{\cos^2\theta}{\cos^2\theta+2}\,,\qquad\qquad \sin^2\theta_{23}=\frac{1}{2}+\frac{\sqrt{6} \sin  2\theta \cos \sigma}{2\cos^2\theta+4}\,,\\
\nonumber \sin \delta_{CP} & = & - \frac{\text{sign}(\sin2\theta)(\cos^2\theta+2) \sin \sigma}{\sqrt{(\cos^2\theta+2)^2-6\sin^2 2\theta \cos^2\sigma}}\,,\quad \cos \delta_{CP}=\frac{\text{sign}(\sin2\theta)(5 \cos^2 \theta-2)  \cos\sigma }{ \sqrt{(\cos^2 \theta +2)^2-6 \cos^2\sigma \sin^2 2 \theta}}\,,\\
\label{eq:mix14}\tan\delta_{CP} & = & -\frac{(\cos^2\theta+2)\tan\sigma}{5\cos^2\theta-2}\,,\qquad
\phi_{12}=\rho + \frac{(k_1-k_2)\pi}{2}\,,\qquad\phi_{13}=\rho+\frac{(k_1-k_3)\pi}{2}\,.
\label{eq:oss-14}
\end{eqnarray}
Note that in the symmetric parametrization the CP violating phase characterizing neutrino oscillations is given by the invariant combination of the fundamental Majorana phases as $\delta_{CP} = \phi_{13} - \phi_{12} - \phi_{23}$~\cite{Rodejohann:2011vc}.  
In Eq.~\eqref{eq:oss-14} we have not explicitly written the phase $\phi_{23}$. One can obtain it easily by inverting the relation $\delta_{CP} = \phi_{13} - \phi_{12} - \phi_{23}$ between $\phi_{23}$, $\delta_{CP}$ and the two other phases. 

\begin{figure}[h!]
 \begin{center}
  \begin{tabular}{cc}
   \includegraphics[width=0.98\linewidth]{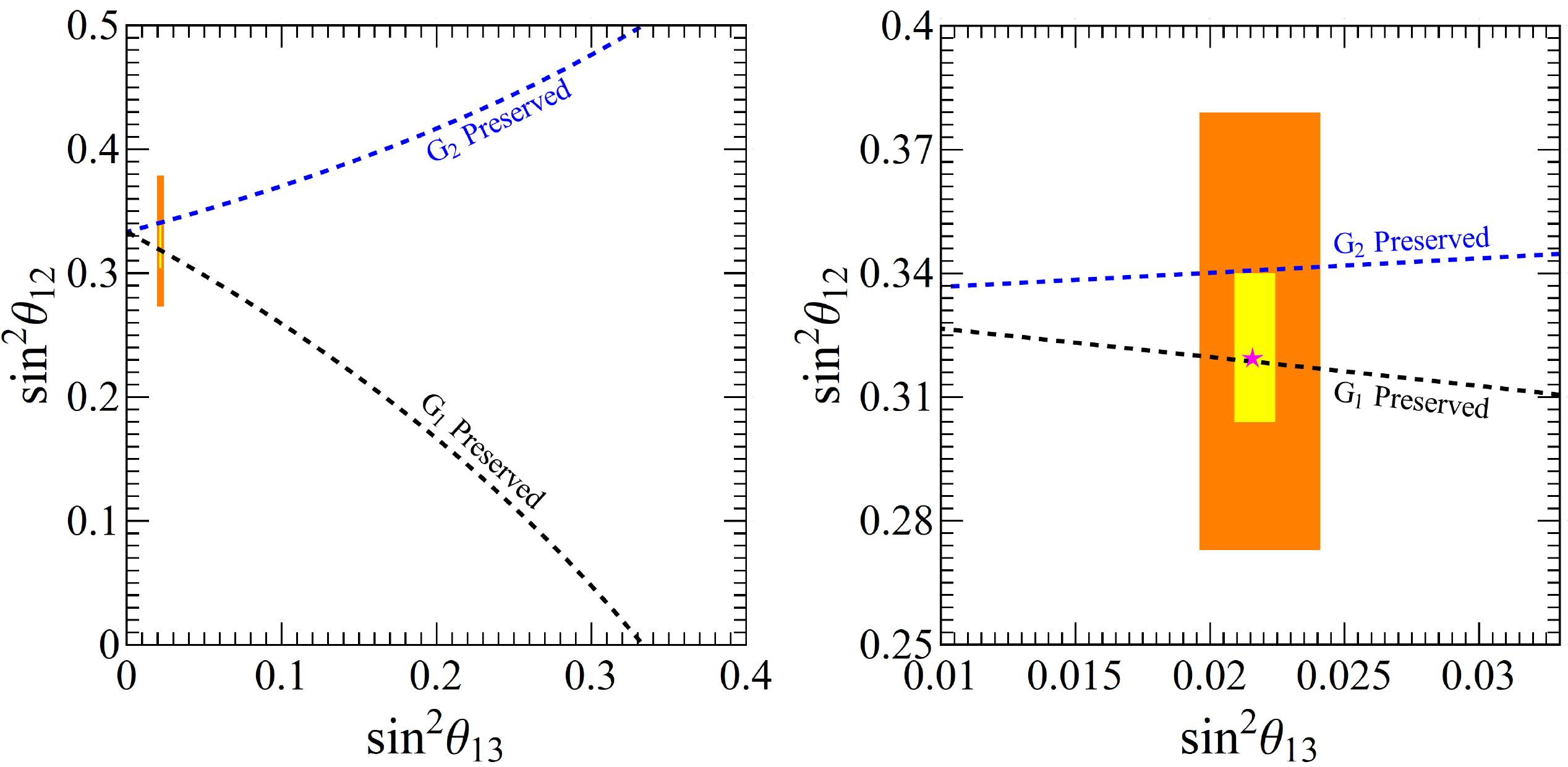}  
  \end{tabular}
 \caption{ Relating solar and reactor angles: predicted correlation between $\sin^2\theta_{12}$ and 
$\sin^2\theta_{13}$. The black dashed line corresponds to the case where 
$G_1$ is preserved by the neutrino sector (Eq.~\eqref{eq:cor1}, left), 
while the blue dashed line refers to the case 
where $G_2$ is preserved by the neutrino sector (Eq.~\eqref{eq:cor2}, left). The right panel is a zoom of the left one.}
\label{fig:g1L}
 \end{center}
 \end{figure}

\begin{figure}[h!]
 \begin{center}
  \begin{tabular}{cc}
   \includegraphics[width=0.98\linewidth]{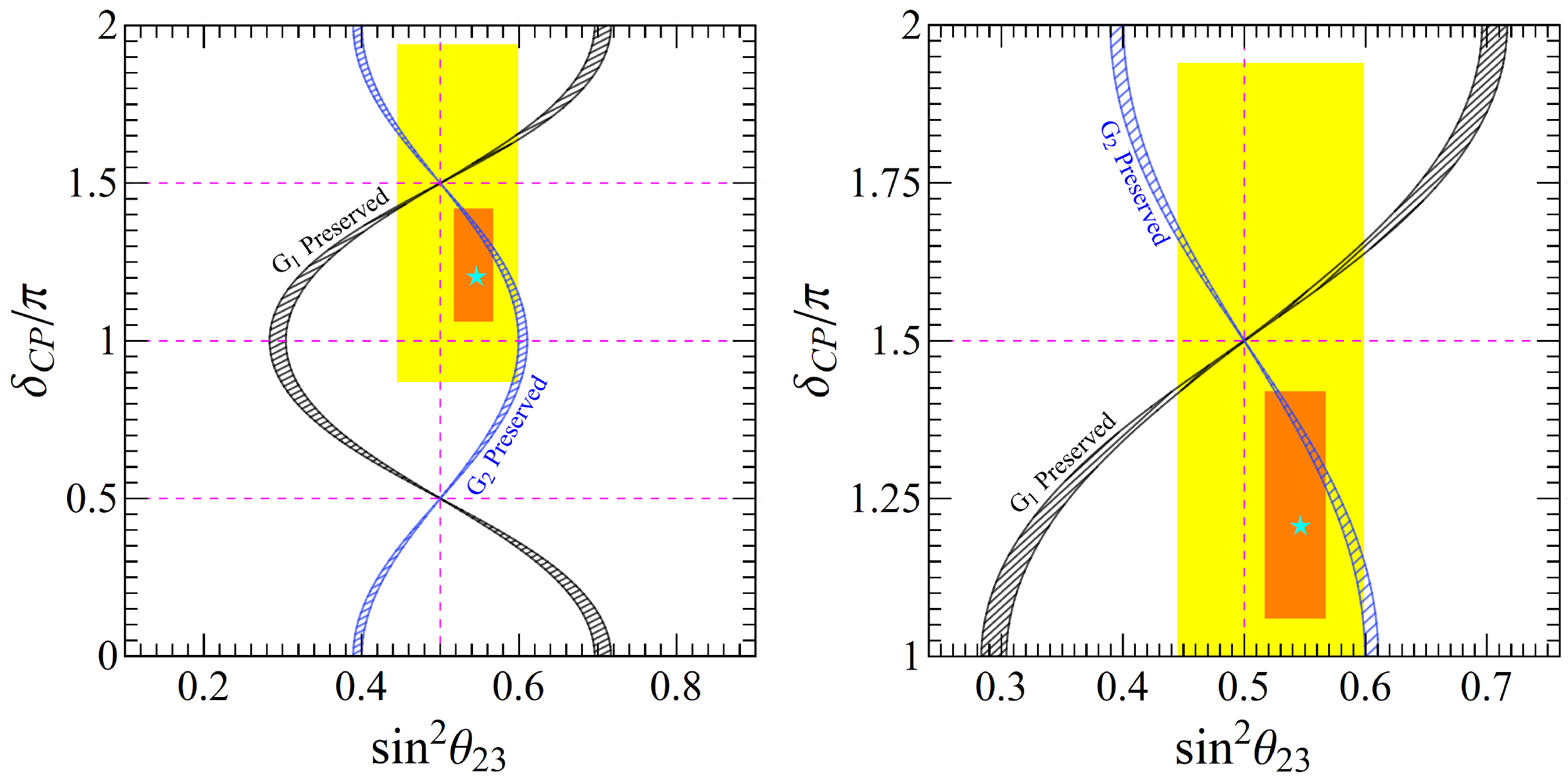}  
  \end{tabular}
 \caption{ Predicted correlation between $\sin^2\theta_{23}$ and $\delta_{CP}$. 
The black region corresponds to the case where $G_1$ is preserved by the neutrino sector, as
given from (Eq.~\eqref{eq:cor1}, right). 
The blue region corresponds to the case where $G_2$ is preserved by the neutrino sector, as given by (Eq.~\eqref{eq:cor2}, right). The right panel is a zoom of the left one.
}
 \label{fig:g1R}
 \end{center}
\end{figure}

Notice that the modulus of entries in the 1st column of $U_{lep}$ in Eq.~\eqref{eq:ux14} is fixed and is 
independent of all parameters. This leads to correlations between the mixing parameters of  $U_{lep}$ 
which are given by
\begin{eqnarray}
\f\cos^2\theta_{12}\cos^2\theta_{13}=\frac{2}{3}\,,\qquad
\tan2\theta_{23}\cos\delta_{CP}=\frac{5\sin^2\theta_{13}-1}{4\tan\theta_{12}\sin\theta_{13}}
=\frac{5\sin^2\theta_{13}-1}{2\sin\theta_{13}\sqrt{2-6\sin^2\theta_{13}}}\,. 
\label{eq:cor1}
\end{eqnarray}
The equation on the left in Eq.~\eqref{eq:cor1} relates the reactor and the solar angle while, for given 
values of these, the right-hand-side equation correlates the CP phase in oscillations to the atmospheric 
angle. These correlations can be used to test the mixing matrix of Eq.~\eqref{eq:ux14} at current and future 
oscillation experiments. We stress that these correlations are a generic feature of mass matrices which 
preserve $G_1$ symmetry. These are displayed in Fig.~\ref{fig:g1L}, while the correlations following from 
the right hand-side equation are shown in Fig.~\ref{fig:g1R}.



\subsubsection*{Real TBM Limit: $\rho, \sigma \to 0$}


So far we have only discussed the consequences arising from the perturbation term preserving 
 the $X_1, X_4$ CP symmetries of the complex TBM matrix. 
In order to obtain the leptonic mixing matrix related with the CP symmetries of the real TBM matrix, one can simply take the limit $\rho, \sigma \to 0$
in Eq.~\eqref{eq:ux14}. Taking this limit we get 
\begin{equation}
\label{eq:rux14}
\left [ U_{lep} \right]_{\rho, \sigma \to 0}  \, = \,
\left(\begin{array}{ccc}
\sqrt{\frac{2}{3}}  & \frac{\cos \theta}{\sqrt{3}} &  \frac{\sin \theta}{\sqrt{3}}
       \\
-\frac{1}{\sqrt{6}} & \frac{\cos \theta}{\sqrt{3}} - \frac{\sin \theta}{\sqrt{2}}
&  \frac{\sin \theta}{\sqrt{3}} + \frac{ \cos \theta}{\sqrt{2}}
       \\
\frac{1}{\sqrt{6}}  & - \frac{\sin \theta}{\sqrt{2}} - \frac{\cos \theta}{\sqrt{3}}
&  \frac{\cos \theta}{\sqrt{2}} - \frac{\sin \theta}{\sqrt{3}}
\end{array}\right)\,Q_{\nu} \,.
\end{equation}
The mixing parameters from Eq.~\eqref{eq:rux14} can also be obtained in straightforward way by 
taking the limit  $\rho, \sigma \to 0$ in Eq.~\eqref{eq:oss-14}. Taking this limit we get 
\begin{eqnarray}
\nonumber\sin^2\theta_{13} & = & \frac{\sin^2\theta}{3}\,,\qquad\qquad 
\sin^2\theta_{12}=\frac{\cos^2\theta}{\cos^2\theta+2}\,,
\qquad\qquad \sin^2\theta_{23}=\frac{1}{2}+\frac{\sqrt{6} \sin  2\theta }{2\cos^2\theta+4}\,,\\
\tan\delta_{CP} & = & 0 \,,\qquad\qquad\qquad
\phi_{12} =  \frac{(k_1-k_2)\pi}{2}\,,\qquad\qquad\qquad 
\phi_{13} = \frac{(k_1-k_3)\pi}{2}\,.
\label{eq:oss-r14}
\end{eqnarray}
By comparing Eq.~\eqref{eq:oss-r14} with Eq.~\eqref{eq:oss-14} one sees that the mixing angles $\theta_{12}$ and $\theta_{13}$ 
remain the same in both cases. However, the range of possible values for $\theta_{23}$ as well as the phases do change.
Notice that in the limit of $\rho, \sigma \to 0$ from Eq.~\eqref{eq:oss-r14} it follows that $\tan\delta_{CP} \to 0$, so that CP 
will be conserved in neutrino oscillations.
Moreover, both Majorana phases become some integer multiples of $\pi/2$ and therefore they correspond to just CP signs~\cite{Schechter:1981hw}.
It follows that the mixing matrix obtained from the real TBM matrix by assuming the $X_1, X_4$ CP symmetries leads necessarily to a CP conserving theory.

Notice however that, since the correlations amongst mixing parameters given in Eq.~\eqref{eq:cor1} 
are independent of $\rho$ and $\sigma$, they remain the same.
However, since for $\rho, \sigma \to 0$ we also have $\cos \delta_{CP} \to 1$, one sees that in this case $\theta_{23}$ 
 gets confined to a narrow range, now ruled out by oscillation data to a very high significance, see Fig~\ref{fig:g1R}.
Thus, the leptonic mixing matrix of Eq.~\eqref{eq:rux14} preserving $X_1, X_4$ CP symmetries of real TBM is ruled out as 
it can not account for atmospheric oscillations.

\subsection{$G_1$ flavor and  $X_2, X_3$ CP symmetries preserved}
\label{sec:cp23-g1}

Now we turn to the second case of two CP symmetries, $X_2, X_3$ this time, which also leads to 
conservation of the $G_1$ flavor symmetry.
As in the previous case, when $X_2$ and $X_3$ CP symmetries are preserved,
the full mass term $ M_\nu = M_\nu^{(0)}+\delta M_\nu$ satisfies
\begin{equation}
\label{eq:hatm23}
U_{cTBM}^T (M_\nu^{(0)} + \delta M_\nu)U_{cTBM} =
\left(\begin{array}{ccc}
 m_1 & 0 & 0 \\
 0 & m_2 & i\delta m \\
 0 & i\delta m & m_3
\end{array}\right)\,.
\end{equation}
As a result of the presence of the perturbation term $\delta M_\nu$ which only preserves ($X_2, X_3$),  
two of the four CP symmetries of Eq.~\eqref{eq:ccp-sym}, the full mass matrix $M_\nu$ is not fully
 diagonalized by the $U_{cTBM}$ matrix.
However, Eq.~\eqref{eq:hatm23} can be easily diagonalized by $\text{diag}(1,-i,1)O_{23}$ with
\begin{equation}
O_{23}=
\left(\begin{array}{ccc}
1 & 0 & 0 \\
0 & \cos\theta & \sin\theta \\
0 & -\sin\theta & \cos\theta
\end{array}\right)\,,
\qquad\text{with}\qquad
\begin{aligned}
&\tan2\theta=\frac{2\delta m}{m_3+m_2}\,,\\
&\cos2\theta=\frac{m_3+m_2}{\sqrt{(m_3+m_2)^2+4\delta m^2}}\,.
\end{aligned}
\end{equation}
The mass eigenvalues are given by
\begin{eqnarray}
m_1^\prime=m_1, \,
m_2^\prime=\frac{1}{2}\big(-m_2+m_3-\sqrt{(m_3+m_2)^2+4\delta m^2}\big), \,
m_3^\prime=\frac{1}{2}\big(-m_2+m_3+\sqrt{(m_3+m_2)^2+4\delta m^2}\big).
\end{eqnarray}
Again, since we are working in the diagonal charged lepton basis, the lepton mixing matrix $U_{lep}$ is given by
\begin{equation}
U_{lep} = U_{cTBM}\,\text{diag}(1,-i,1)\,O_{23}\,Q_{\nu} \, = \,
\left(\begin{array}{ccc}
\sqrt{\frac{2}{3}}
& - \frac{i e^{-i \rho} \cos \theta}{\sqrt{3}}
& - \frac{i e^{-i \rho} \sin \theta}{\sqrt{3}}
       \\
-\frac{e^{i \rho}}{\sqrt{6}}
& - \frac{i \cos \theta}{\sqrt{3}} - \frac{e^{-i \sigma} \sin \theta}{\sqrt{2}}
& - \frac{i \sin \theta}{\sqrt{3}} + \frac{e^{-i \sigma} \cos \theta}{\sqrt{2}}
       \\
\frac{e^{i (\rho + \sigma)}}{\sqrt{6}}
& - \frac{\sin \theta}{\sqrt{2}} + \frac{i e^{i \sigma} \cos \theta}{\sqrt{3}}
&  \frac{\cos \theta}{\sqrt{2}} + \frac{i e^{i \sigma} \sin \theta}{\sqrt{3}}
\end{array}\right)\,Q_{\nu} \,.
\label{eq:ux23}
\end{equation}
As before, here we have
 $Q_{\nu} = \text{diag} (e^{i k_1 \pi /2}, e^{i k_2 \pi /2} , e^{i k_3 \pi /2} )$ is a diagonal matrix of phases.
This mixing matrix is nothing but the gTBM matrix discussed recently in \cite{Chen:2018eou}, fixing the choice
 $Q_{\nu} = \text{diag} (1,i,1)$.

From the mixing matrix Eq.~\eqref{eq:ux23} one can extract the mixing parameters in the symmetric 
parametrization. They are given by,
\begin{eqnarray}
\nonumber\sin^2\theta_{13} & = & \frac{\sin^2\theta}{3}\,,
\qquad\qquad \sin^2\theta_{12}=\frac{\cos^2\theta}{\cos^2\theta+2}\,,
\qquad\qquad  \sin^2\theta_{23}=\frac{1}{2}+\frac{\sqrt{6}\sin  2\theta\sin\sigma}{2\cos^2\theta+4}\,,\\
\nonumber \sin \delta_{CP} & = & \frac{\text{sign}(\sin2\theta)(\cos^2\theta+2) \cos\sigma}{\sqrt{(\cos^2\theta+2)^2-6\sin^2 2\theta \sin^2\sigma}}\,,
\quad \cos \delta_{CP}=\frac{\text{sign}(\sin2\theta)(5\cos^2\theta-2)\sin\sigma}{\sqrt{(\cos^2 \theta +2)^2-6\sin^22\theta\sin^2\sigma}}\,,\\
\tan\delta_{CP} & = & \frac{(\cos^2\theta+2)\cot\sigma}{5\cos^2\theta-2}\,,\qquad
\phi_{12}=\rho + \frac{(k_1-k_2+1)\pi}{2}\,,
\qquad\phi_{13}=\rho + \frac{(k_1-k_3+1)\pi}{2}\,.
\label{eq:mix23}
\end{eqnarray}

The expression for the phase $\phi_{23}$ can be extracted by inverting its relation with $\delta_{CP}$ and 
the other phases. The formula is lengthy, so we do not write it explicitly.
From the above results one sees that, owing to the fact that both $(X_1\,,\,X_4)$ and $(X_2\,,\,X_3)$ cases 
preserve the same flavuor symmetry $G_1$, they lead to the same correlations between mixing parameters given in Eq.\eqref{eq:cor1}.
In particular, since the same flavour symmetry $G_1$ is conserved in both cases, Eq.~\eqref{eq:mix23} can be obtained from Eq.~\eqref{eq:mix14} just by 
redefining 
\begin{eqnarray}
\rho \to \rho + \pi/2 \mathrm{~~~and~~~} \sigma \to \sigma-\pi/2.
\end{eqnarray}


\subsubsection*{Real TBM Limit: $\rho, \sigma \to 0$}


Again, as before, the mixing matrix corresponding to conserved $X_2, X_3$ CP symmetries of the real 
TBM matrix Eq.~\eqref{eq:tbm} can be obtained from Eq.~\eqref{eq:ux23} by simply taking the limit 
$\rho, \sigma \to 0$. Its form is given by
\begin{equation}
\label{eq:rux23}
\left [ U_{lep} \right]_{\rho, \sigma \to 0}  \, = \,
\left(\begin{array}{ccc}
\sqrt{\frac{2}{3}}  & - \frac{i \cos \theta}{\sqrt{3}} & - \frac{i \sin \theta}{\sqrt{3}}
       \\
-\frac{1}{\sqrt{6}} & - \frac{i \cos \theta}{\sqrt{3}} - \frac{\sin \theta}{\sqrt{2}}
& - \frac{i \sin \theta}{\sqrt{3}} + \frac{\cos \theta}{\sqrt{2}}
       \\
\frac{1}{\sqrt{6}}  & - \frac{\sin \theta}{\sqrt{2}} + \frac{i \cos \theta}{\sqrt{3}}
&  \frac{\cos \theta}{\sqrt{2}} + \frac{i \sin \theta}{\sqrt{3}}
\end{array}\right)\,Q_{\nu} \,.
\end{equation}
The mixing parameters corresponding to Eq.~\eqref{eq:rux23} are readily
obtained from Eq.~\eqref{eq:mix23} by taking $\rho, \sigma \to 0$. They are given as
\begin{eqnarray}
\nonumber\sin^2\theta_{13} & = & \frac{\sin^2\theta}{3}\,,
\qquad\qquad \sin^2\theta_{12} = \frac{\cos^2\theta}{\cos^2\theta+2}\,,
\qquad\qquad  \sin^2\theta_{23} = \frac{1}{2}\,,\\
\delta_{CP} & = & \pm \frac{\pi}{2}\,,
\qquad \qquad \quad \phi_{12} =  \frac{(k_1-k_2+1)\pi}{2}\,,
\qquad \quad \phi_{13} = \frac{(k_1-k_3+1)\pi}{2}\,.
\label{eq:rmix23}
\end{eqnarray}
One sees from Eq.~\eqref{eq:rmix23} that in this case both the atmospheric angle $\theta_{23}$ as well as the Dirac CP phase are predicted to be maximal.
We remind the reader that the $X_3$ CP symmetry of the real TBM matrix (see Eq.~\eqref{eq:rcp-sym}) is nothing but the well-known $\mu-\tau$ symmetry. 
The prediction of maximal atmospheric mixing and maximal Dirac CP phase in this case appears as a natural consequence of it.
Before moving on we note also that, for the choice of $Q_{\nu} = \text{diag} (1,i,1)$, the leptonic mixing matrix of Eq.~\eqref{eq:rux23} has already been 
discussed in the literature, for example in \cite{Feruglio:2012cw,Li:2013jya,Rodejohann:2017lre}. Moreover, this matrix has been explored as one of the 
limiting cases of the gTBM matrix of \cite{Chen:2018eou}.

\subsection{$G_2$ flavor and $X_2,X_4$ CP symmetries preserved}
\label{sec:cp24-g2}

Now we move on to the cases when the $G_2$ flavor symmetry is preserved. Just like in the previous case, here also two different combinations of two CP symmetries, namely $(X_2, X_4)$ and $(X_1, X_3)$ preserve the $G_2$ flavor symmetry.
We first consider the case when the perturbation term preserves $X_2$ and $X_4$ CP symmetries.
In this case the perturbation term satisfies
\begin{eqnarray}
X_i^T \delta M_\nu X_i=\delta M_\nu^{\ast}\,,\qquad
\hat{d}_i \, U_{cTBM}^T\delta M_\nu  U_{cTBM}\, \hat{d}_i = (U_{cTBM}^T\delta M_\nu  U_{cTBM})^\ast\,.
\end{eqnarray}
where $i=2, 4$. Thus, $U_{cTBM}^T\delta M_\nu  U_{cTBM}$ must be the form 
\begin{equation}
U_{cTBM}^T \delta M_\nu U_{cTBM} =
\left(\begin{array}{ccc}
 \delta m_1^\prime & 0 & \delta m \\
 0 & \delta m_2^\prime & 0 \\
 \delta m & 0 & \delta m_3^\prime
\end{array}\right)\,.
\label{per-24}
\end{equation}
Again as before $\delta m_1^\prime$, $\delta m_2^\prime$ and $\delta m_3^\prime$ can be
absorbed by $m_1$, $m_2$ and $m_3$.
Thus without loss of generality we can take $\delta m_1^\prime = \delta m_2^\prime = \delta m_3^\prime = 0$ in Eq.~\eqref{per-24} and obtain
\begin{equation}
\delta M_\nu =
U_{cTBM}^T\left(\begin{array}{ccc}
 0 & 0 & \delta m \\
 0 & 0 & 0 \\
 \delta m & 0 & 0
 \end{array}\right)\,.
\end{equation}
Thus the full mass matrix $M_\nu = M_\nu^{(0)}+\delta M_\nu$ satisfies
\begin{equation}
\label{eq:hatm13}
U_{cTBM}^T (M_\nu^{(0)} + \delta M_\nu)U_{cTBM} =
\left(\begin{array}{ccc}
 m_1 & 0 & \delta m \\
 0 & m_2 & 0 \\
 \delta m & 0 & m_3
\end{array}\right)\,.
\end{equation}
As before, owing to the presence of perturbation term $\delta M_\nu$, the mixing matrix $U_{cTBM}$ 
does not fully diagonalize the full mass matrix $M_\nu$. However Eq.~\eqref{eq:hatm13} can be diagonalized by
\begin{equation}
O_{13}=
\left(\begin{array}{ccc}
\cos\theta & 0 & \sin\theta \\
0 & 1 & 0 \\
-\sin\theta & 0 & \cos\theta
\end{array}\right)\,,
\qquad\text{with}\qquad\quad
\begin{aligned}
&\tan2\theta=\frac{2\delta m}{m_3-m_1}\,,\\
&\cos2\theta=\frac{m_3-m_2}{\sqrt{(m_3-m_1)^2+4\delta m^2}}\,.
\end{aligned}
\end{equation}
The masses are given by
\begin{eqnarray}
m_1^\prime=\frac{1}{2}\big(m_1+m_3-\sqrt{(m_3-m_1)^2+4\delta m^2}\big), \,
m_2^\prime = m_2 , \,
m_3^\prime=\frac{1}{2}\big(m_1+m_3+\sqrt{(m_3-m_1)^2+4\delta m^2}\big)\,.
\end{eqnarray}
Thus the full leptonic mixing matrix in this case is given by
\begin{equation}
U_{lep}=U_{cTBM}\,O_{13}\,Q_{\nu}\, = \,
\left(\begin{array}{ccc}
\sqrt{\frac{2}{3}} \cos \theta
& \frac{e^{-i \rho}}{\sqrt{3}}
& \sqrt{\frac{2}{3}} \sin \theta
       \\
- \frac{e^{i \rho} \cos \theta}{\sqrt{6}} - \frac{e^{-i \sigma} \sin \theta}{\sqrt{2}}
& \frac{1}{\sqrt{3}}
& - \frac{e^{i \rho} \sin \theta}{\sqrt{6}} + \frac{e^{-i \sigma} \cos \theta}{\sqrt{2}}
       \\
\frac{e^{i (\rho + \sigma)} \cos \theta}{\sqrt{6}} - \frac{ \sin \theta}{\sqrt{2}}
& - \frac{e^{i \sigma}}{\sqrt{3}}
& \frac{e^{i (\rho + \sigma)} \sin \theta}{\sqrt{6}} + \frac{ \cos \theta}{\sqrt{2}}
\end{array}\right)\,Q_{\nu} \,.
\label{eq:ux24}
\end{equation}
where, as before, $Q_{\nu} = \text{diag} (e^{i k_1 \pi /2}, e^{i k_2 \pi /2} , e^{i k_3 \pi /2} )$ is a diagonal 
matrix of phases. 

The mixing parameters associated to the mixing matrix following from Eq.~\eqref{eq:ux24} are given by
\begin{eqnarray}
\nonumber\sin^2\theta_{13} & = & \frac{2 \sin^2\theta}{3}\,,
\qquad\sin^2\theta_{12}=\frac{1}{2\cos^2\theta+1}\,,
\qquad\sin^2\theta_{23}=\frac{1}{2}-\frac{\sqrt{3} \sin  2\theta \cos(\rho + \sigma)}{4\cos^2\theta+2}\,,\\
\nonumber\sin \delta_{CP} & = & -\frac{\text{sign}(\sin2\theta) (2\cos^2\theta+1)\sin(\rho + \sigma)}{\sqrt{(2\cos^2\theta+1)^2-3 \cos^2(\rho + \sigma) \sin^2 2\theta}}\,,
\qquad \cos \delta_{CP}=\frac{\text{sign}(\sin2\theta) (4 \cos^2 \theta-1) \cos(\rho + \sigma)}{ \sqrt{(2\cos^2 \theta+1)^2-3 \cos^2(\rho + \sigma) \sin^2 2 \theta}}\,,\\
\label{eq:mix24}\tan\delta_{CP} & = & -\frac{(2\cos^2\theta+1)\tan(\rho + \sigma)}{4\cos^2\theta-1}\,,\qquad\qquad
\phi_{12}=\rho + \frac{(k_1-k_2)\pi}{2}\,,
\qquad\phi_{13}=\frac{(k_1-k_3)\pi}{2}\,.
\label{eq:mix-ux24}
\end{eqnarray}
Again as before, the expression for $\phi_{23}$ can also be readily obtained from Eq.~\eqref{eq:mix-ux24} 
 using the relation between $\phi_{23}$ and other phases. 
From Eq.~\eqref{eq:mix-ux24} we find that the mixing parameters are again correlated with each other. 
The correlations are given by
\begin{eqnarray}
\sin^2\theta_{12}\cos^2\theta_{13}=\frac{1}{3}\,,
\qquad \tan2\theta_{23} \cos\delta_{CP}=\frac{\cos2\theta_{13}\tan\theta_{12}}{\sin\theta_{13}}
=\frac{\cos2\theta_{13}}{\sin\theta_{13}\sqrt{2-3\sin^2\theta_{13}}}\,.
\label{eq:cor2}
\end{eqnarray}
These correlations lead to strong predictions for the oscillations parameters as shown by the blue curves in Fig. \ref{fig:g1L} and Fig. \ref{fig:g1R}.
Notice that although the correlations are again between the same oscillation parameters i.e. one between $\theta_{12} - \theta_{13}$ angles and the other between
$\theta_{23} - \delta_{CP}$ but the form of correlations is very different from the obtained in Eq.~\eqref{eq:cor1} for the two cases of $G_1$ flavor symmetry.
Thus these correlations and their associated predictions can be used to distinguish between the $G_1$ and $G_2$ flavor symmetries, as can be seen from Fig. \ref{fig:g1L} and Fig. \ref{fig:g1R}. 


\subsubsection*{Real TBM Limit: $\rho, \sigma \to 0$}


Again, as in previous cases, the mixing matrix corresponding to the conserved $X_2, X_4$ CP 
symmetries of the real TBM matrix Eq.~\eqref{eq:tbm} can be obtained from Eq.~\eqref{eq:ux24} by taking the 
limit $\rho, \sigma \to 0$. The results is  
\begin{equation}
\label{eq:rux24}
\left [ U_{lep} \right]_{\rho, \sigma \to 0}  \, = \,
\left(\begin{array}{ccc}
\sqrt{\frac{2}{3}} \cos \theta
& \frac{1}{\sqrt{3}}
& \sqrt{\frac{2}{3}} \sin \theta
       \\
- \frac{\cos \theta}{\sqrt{6}} - \frac{\sin \theta}{\sqrt{2}}
& \frac{1}{\sqrt{3}}
& - \frac{\sin \theta}{\sqrt{6}} + \frac{\cos \theta}{\sqrt{2}}
       \\
\frac{\cos \theta}{\sqrt{6}} - \frac{ \sin \theta}{\sqrt{2}}
& - \frac{1}{\sqrt{3}}
& \frac{\sin \theta}{\sqrt{6}} + \frac{ \cos \theta}{\sqrt{2}}
\end{array}\right)\,Q_{\nu} \,.
\end{equation}
The mixing parameters corresponding to Eq.~\eqref{eq:rux24} can be obtained from Eq.~\eqref{eq:mix-ux24} 
by taking $\rho, \sigma \to 0$ and are given by 
\begin{eqnarray}
\nonumber\sin^2\theta_{13} & = & \frac{2 \sin^2\theta}{3}\,,
\qquad\sin^2\theta_{12}=\frac{1}{2\cos^2\theta+1}\,,
\qquad\sin^2\theta_{23}=\frac{1}{2}-\frac{\sqrt{3} \sin  2\theta }{4\cos^2\theta+2}\,,\\
\tan\delta_{CP} & = & 0\,,\qquad\qquad\qquad
\phi_{12} = \frac{(k_1-k_2)\pi}{2}\,,
\qquad \qquad \phi_{13} = \frac{(k_1-k_3)\pi}{2}\,.
\label{eq:mix-rux24}
\end{eqnarray}

One sees from Eq.~\eqref{eq:mix-rux24} that not only the Dirac phase vanishes, but also the Majorana phases 
$\phi_{12}, \phi_{13}$ take on CP-conserving values, since they are integer multiples of $\pi/2$, corresponding 
to Majorana CP signs~\cite{Schechter:1981hw}. 
Thus, just like the real TBM limit of the first case of $G_1$ flavor symmetry discussed in Section \ref{sec:cp14-g1}, 
here too the $(X_2, X_4)$ preserving case of real TBM CP symmetries predicts no CP violation.
As before, since the correlations between oscillation parameters of Eq.~\eqref{eq:cor2} are $\rho$ and $\sigma$ independent,
they remain the same for the real TBM case as well. 
However, since now $\delta_{CP} \to 0$, the angle $\theta_{23}$ gets confined to a narrow range, as can be seen from Fig. \ref{fig:g1R}.
Particularly for $\delta_{CP} = \pi$, the predicted range of the atmospheric angle $\theta_{23}$ lies at the edge of currently allowed 
3$\sigma$ range~\cite{deSalas:2017kay}, and may be ruled out in the near future.

\subsection{$G_2$ flavor and $X_1,X_3$ CP symmetries preserved}
\label{sec:cp13-g2}

The other option for two CP symmetries that preserves the $G_2$ flavor symmetry is the case 
 where the $X_1$ and $X_3$ CP symmetries are preserved.
 In this case, as before the leading term $M_\nu^{(0)}$ of the neutrino mass matrix preserves all 
four CP symmetries in Eq.~\eqref{eq:ccp-sym}, while the perturbation term $\delta M_\nu$ only preserves 
the $X_1$ and $X_3$ CP symmetries. Therefore in this case we have
\begin{equation}
U_{cTBM}^T ( M_\nu^{(0)} + \delta M_\nu ) U_{cTBM} =
\left(\begin{array}{ccc}
 m_1 & 0 & i\delta m \\
 0 & m_2 & 0 \\
 i\delta m & 0 & m_3
\end{array}\right)\,,
\end{equation}
This can be diagonalized by $\text{diag}(-i,1,1) O_{13}$ where
\begin{equation}
O_{13}=
\left(\begin{array}{ccc}
\cos\theta & 0 & \sin\theta \\
0 & 1 & 0 \\
-\sin\theta & 0 & \cos\theta
\end{array}\right)\,,
\qquad\text{with}\qquad\quad
\begin{aligned}
&\tan2\theta=\frac{2\delta m}{m_3+m_1}\,,\\
&\cos2\theta=\frac{m_3-m_2}{\sqrt{(m_3+m_1)^2+4\delta m^2}}\,.
\end{aligned}
\end{equation}
The masses in this case are given by
\begin{eqnarray}
m_1^\prime = \frac{1}{2}\big(-m_1+m_3-\sqrt{(m_3+m_1)^2+4\delta m^2}\big),\,
m_2^\prime = m_2 , \,
m_3^\prime=\frac{1}{2}\big(-m_1+m_3+\sqrt{(m_3+m_1)^2+4\delta m^2}\big).
\end{eqnarray}

The leptonic mixing matrix in this case is given as
\begin{equation}
U_{lep} = U_{cTBM}\,\text{diag}(-i,1,1)\,O_{13}\,Q_{\nu} \, = \,
\left(\begin{array}{ccc}
-i \sqrt{\frac{2}{3}} \cos \theta
& \frac{e^{-i \rho}}{\sqrt{3}}
& -i \sqrt{\frac{2}{3}} \sin \theta
       \\
\frac{i e^{i \rho} \cos \theta}{\sqrt{6}} - \frac{e^{-i \sigma} \sin \theta}{\sqrt{2}}
& \frac{1}{\sqrt{3}}
&  \frac{i e^{i \rho} \sin \theta}{\sqrt{6}} + \frac{e^{-i \sigma} \cos \theta}{\sqrt{2}}
       \\
- \frac{i e^{i (\rho + \sigma)} \cos \theta}{\sqrt{6}} - \frac{ \sin \theta}{\sqrt{2}}
& - \frac{e^{i \sigma}}{\sqrt{3}}
& - \frac{i e^{i (\rho + \sigma)} \sin \theta}{\sqrt{6}} + \frac{ \cos \theta}{\sqrt{2}}
\end{array}\right)\,Q_{\nu} \,.
\label{eq:ux13}
\end{equation}
We recall that $Q_{\nu} = \text{diag} (e^{i k_1 \pi /2}, e^{i k_2 \pi /2} , e^{i k_3 \pi /2} )$ is again a diagonal matrix of phases.

The mixing parameters can be easily extracted from Eq.~\eqref{eq:ux13} and are given by
\begin{eqnarray}
\sin^2\theta_{13} & = & \frac{2 \sin^2\theta}{3}\,,\qquad \sin^2\theta_{12}=\frac{1}{2\cos^2\theta+1}\,,\qquad \sin^2\theta_{23}=\frac{1}{2}-\frac{\sqrt{3} \sin  2\theta \sin (\rho + \sigma)}{4\cos^2\theta+2}\,, \nonumber \\
\sin \delta_{CP} & = & \frac{\text{sign}(\sin2\theta) (2\cos^2\theta+1) \cos (\rho + \sigma)}{\sqrt{(2\cos^2\theta+1)^2-3\sin^2 2\theta \sin^2(\rho + \sigma)}}\,,
\qquad\cos \delta_{CP}=\frac{\text{sign}(\sin2\theta)(4\cos^2\theta-1) \sin (\rho + \sigma)}{\sqrt{(2\cos^2 \theta+1)^2-3 \sin^2 2 \theta \sin^2(\rho + \sigma)}}\,,\nonumber\\
\tan\delta_{CP} & = & \frac{(2\cos^2\theta+1)\cot(\rho + \sigma)}{4\cos^2\theta-1}\,,\qquad
\phi_{12}=\rho + \frac{(k_1-k_2+1)\pi}{2}\,,
\qquad\phi_{13}=\frac{(k_1-k_3)\pi}{2}\,.
\label{eq:oss-ux13}
\end{eqnarray}
As before $\phi_{23}$ phase can also be obtained from Eq.~\eqref{eq:oss-ux13} in a straightforward way.

Notice the similarities and differences between the two $G_2$ flavor symmetry conserving cases 
of Sections~\ref{sec:cp24-g2} and~\ref{sec:cp13-g2}, associated to conservation of 
$(X_2\,,\,X_4)$ and $(X_1\,,\,X_3)$, respectively. They lead to different mixing angles and 
phases, as can be seen from Eqs.~\eqref{eq:mix-ux24} and \eqref{eq:oss-ux13}, respectively.
However they both still satisfy the same correlations between the oscillation parameters, 
given by Eq.~\eqref{eq:cor2}.
In particular, notice that one can obtain Eq.~\eqref{eq:oss-ux13} from Eq.~\eqref{eq:mix-ux24} by 
redefining 
\begin{eqnarray}
\rho \to \rho + \pi/2 \mathrm{~~~and~~~} \sigma \to \sigma-\pi.%
\end{eqnarray}
The predictions for the neutrino oscillation parameters originating from the two correlations of
 Eq.~\eqref{eq:cor2} are shown in Fig.~\ref{fig:g1L} and Fig.~\ref{fig:g1R} respectively.
The $(X_2\,,\,X_4)$ and $(X_1\,,\,X_3)$ cases also differ in their predictions for the Majorana 
phases as can be Eqs.~\eqref{eq:mix-ux24} and \eqref{eq:oss-ux13}.

Before moving on we would like to highlight the differences between the $G_1$ and $G_2$ flavor symmetries.  
In these two scenarios, not only the expressions for the mixing parameters in terms of the model parameters are very different, 
also the correlations between the physical oscillation parameters, as can be seen from Eqs.~\eqref{eq:cor1} and \eqref{eq:cor2}.
These predicted correlations between neutrino oscillation parameters are shown in Fig.~\ref{fig:g1L} and Fig.~\ref{fig:g1R}. 
One should note the two branches, corresponding to the $G_1$ and $G_2$ flavor symmetries.  
This difference in the predicted correlations can be exploited as a test at upcoming neutrino oscillation experiments.


\subsubsection*{Real TBM Limit: $\rho, \sigma \to 0$}


Just as in previous cases, here we can also get the mixing matrix corresponding 
to the $X_1, X_3$ CP symmetries of the real TBM matrix by taking the limit $\rho, \sigma \to 0$ in Eq.~\eqref{eq:ux13}, leading to
\begin{equation}
\left [ U_{lep} \right]_{\rho, \sigma \to 0}  \, = \,
\left(\begin{array}{ccc}
-i \sqrt{\frac{2}{3}} \cos \theta
& \frac{1}{\sqrt{3}}
& -i \sqrt{\frac{2}{3}} \sin \theta
       \\
\frac{i \cos \theta}{\sqrt{6}} - \frac{\sin \theta}{\sqrt{2}}
& \frac{1}{\sqrt{3}}
&  \frac{i \sin \theta}{\sqrt{6}} + \frac{\cos \theta}{\sqrt{2}}
       \\
- \frac{i \cos \theta}{\sqrt{6}} - \frac{ \sin \theta}{\sqrt{2}}
& - \frac{1}{\sqrt{3}}
& - \frac{i \sin \theta}{\sqrt{6}} + \frac{ \cos \theta}{\sqrt{2}}
\end{array}\right)\,Q_{\nu} \,.
\label{eq:rux13}
\end{equation}

The mixing parameters corresponding to Eq.~\eqref{eq:rux13} can again be obtained from Eq.~\eqref{eq:oss-ux13} by taking the limit $\rho, \sigma \to 0$. 
They are given by 
\begin{eqnarray}
\sin^2\theta_{13} & = & \frac{2 \sin^2\theta}{3} \,,
\qquad \qquad \sin^2\theta_{12} = \frac{1}{2\cos^2\theta+1} \,,
\qquad \qquad \sin^2\theta_{23} = \frac{1}{2} \,,\\
\delta_{CP} & = & \pm \frac{\pi}{2}\,,
\qquad \qquad \qquad \phi_{12} =  \frac{(k_1-k_2+1)\pi}{2}\,,
\qquad \qquad \phi_{13} = \frac{(k_1-k_3)\pi}{2}\,.
\label{eq:oss-rux13}
\end{eqnarray}

One sees how, starting from the real TBM matrix, and using the $X_1, X_3$ CP symmetries, one is lead to maximal Dirac CP phase. 
Using the fact that the oscillation parameter correlations are $\rho, \sigma$ independent, one sees that for this case the 
atmospheric mixing angle $\theta_{23}$ is also maximal, as shown in Fig. \ref{fig:g1R}.
Once again, the prediction of maximal $\delta_{CP}$ and maximal $\theta_{23}$ is natural, since this case preserves the $X_3$ CP symmetry of the real TBM matrix (see Eq.~\eqref{eq:rcp-sym}), which is nothing but the $\mu - \tau$ symmetry. We note that the lepton mixing pattern in Eq.~\eqref{eq:rux13} has been discussed in~\cite{Feruglio:2012cw,Ding:2013hpa}.


\section{Neutrino mass matrix conserving one CP symmetry}
\label{sec:one-cp}


We will now study the case when just one CP symmetry is preserved in the neutrino sector. For simplicity we will assume $\rho = \sigma = 0$, although the generalization to non-zero values of the Majorana phases is trivial. Therefore, our starting point will be the real TBM matrix of Eq.~\eqref{eq:tbm}. The 4 CP symmetries compatible with the real TBM mixing matrix are given by Eq.~\eqref{eq:rcp-sym}.
Note that $X_4$ is just the identity, while $X_3$ is the mu-tau reflection symmetry, which has been extensively studied in the literature \cite{harrison:2002er, Grimus:2003yn}.
Therefore, we will just discuss the cases $X_1$ and $X_2$. 
When a single CP symmetry $X_i = U_{rTBM} \hat{d}_i U^T_{rTBM}$ is preserved in the neutrino sector,  we can combine Eqs.~\eqref{eq:cpmaj} and \eqref{eq:XCons} to obtain 

\begin{eqnarray}
\label{eq:oct9a} d_i(U^T_{rTBM} m_\nu U_{rTBM})d_i = (U^T_{rTBM} m_\nu U_{rTBM})^\ast\,.
\end{eqnarray}

Consequently the matrix form of $U^T_{rTBM} m_\nu U_{rTBM}$ is given by 

\begin{eqnarray}
U^T_{rTBM} m_\nu U_{rTBM} =
\left ( \begin{matrix}
m_1 & \delta m_{12} &\delta m_{13} \cr
\delta m_{12} & m_2 &\delta m_{23} \cr
\delta m_{13} &\delta m_{23} & m_3
\end{matrix} \right )\,.
\end{eqnarray}
where $m_1$, $m_2$ and $m_3$ are real and $\delta m_{12}$, $\delta m_{13}$ and $\delta m_{23}$ are either pure real or pure imaginary. For example, when $i=1$, $d_1=\text{diag}(1,-1,-1)$, $\delta m_{12}$, $\delta m_{13}$ are pure imaginary and $\delta m_{23}$ is real. When $i=4$ and $d_4=\text{diag}(1,1,1)$, $\delta m_{12}$, $\delta m_{13}$ and $\delta m_{23}$ are real. One can split the matrix $m_\nu$ in terms of the mass parameters as 

\begin{eqnarray}
\nonumber
m_\nu&=&
\frac{m_1}{6} \left(
\begin{matrix}
 4 & -2 & 2 \cr
 -2 & 1 & -1 \cr
 2  & -1 & 1
\end{matrix}
\right)
+\frac{m_2}{3}
\left(
\begin{matrix}
 1 & 1 & -1 \cr
 1 & 1 & -1 \cr
 -1 & -1 & 1
\end{matrix}
\right)
+\frac{m_3}{2}
\left(
\begin{matrix}
 0 & 0 & 0 \cr
 0 & 1 & 1 \cr
 0 & 1 & 1
\end{matrix}
\right)\\
&&+\frac{\delta m_{12}}{3\sqrt{2}}
\left(\begin{matrix}
 4 & 1 & -1 \cr
 1 & -2 & 2 \cr
 -1 & 2 & -2
\end{matrix}\right)
+\frac{\delta m_{13}}{\sqrt{3}}
\left(\begin{matrix}
 0 & 1 & 1 \cr
 1 & -1 & 0 \cr
 1 & 0 & 1
\end{matrix}\right)
+\frac{\delta m_{23}}{\sqrt{6}}
\left(\begin{matrix}
 0 &  1 & 1 \cr
 1 & 2 & 0 \cr
 1 & 0 & -2
\end{matrix}\right)\,.
\end{eqnarray}

From Eq.~\eqref{eq:oct9a} one can see that $d_i^{1/2} U^T_{rTBM} m_\nu U_{rTBM} d_i^{1/2}$  is a real matrix which can be diagonalized by a 3-dimensional orthogonal matrix $O_{3\times 3}$, see Appendix~\ref{sec:o33} for details. Therefore the neutrino mass matrix $m_\nu$ is diagonalized by the following unitary transformation

\begin{eqnarray}
 U = {U_{rTBM}}d^{1/2}_i O_{3\times 3} Q_{\nu}\,.
 \label{generalonecpneutrino}
\end{eqnarray}
where the $Q_\nu$ matrix is given as
\begin{equation}
Q_\nu=
\left( \begin{array}{ccc}
\!\!  e^{ik_1\pi/2} \!&\!  0 \!&\! 0 \!\!\\
\!\!  0~ \!&\! e^{ik_2\pi/2}  \!&\! 0 \!\!\\
\!\!  0~ \!&\! 0 \!&\! e^{ik_3\pi/2} \!\!
 \end{array} \right)\,.
\end{equation}
where $k_i$ take on integer values. Notice that $U_{rTBM}$ (real TBM matrix) and not $U_{cTBM}$ (complex version of the TBM matrix) appears in Eq.~\eqref{generalonecpneutrino}. This is due to the fact that we have chosen $\sigma=\rho = 0$, for simplicity. We will not consider the cases $X_3$ and $X_4$ since they are trivial, as discussed before. 
%

\subsection{Preserving the $X_1$ CP symmetry}

This is the first non-trivial case. Following Eq.~\eqref{generalonecpneutrino}, when $i=1$ the lepton mixing matrix is given by 

\begin{equation}
U_{lep} = U_{rTBM}\text{diag}(1, i, i) O_{3\times3} Q_{\nu}\,.
\label{UX1}
\end{equation}

The mixing parameters of such mixing matrix are
\begin{eqnarray}
\sin^2 \theta_{13}  & = & \frac{1}{3} \left(\cos^2 \theta_2 \sin^2\theta_1+2 \sin^2\theta_2\right)\,, \nonumber \\  
\sin^2\theta_{12}&=&\frac{(\cos\theta_1\cos\theta_3-\sin\theta_1\sin\theta_2\sin\theta_3)^2+2\cos^2\theta_2\sin^2\theta_3}{(1+\cos^2\theta_1)\cos^2\theta_2+1}\,,\nonumber \\
\sin^2 \theta_{23}  & = &  \frac{1}{2} + \frac{2 \sqrt{6} \cos^2\theta_2\sin2\theta_1}{\cos 2\theta_1+2\left(\cos^2\theta_1+1\right) \cos 2\theta_2+7}\,, \nonumber \\
J_{CP} & = & -\frac{\left(3-5\cos^2\theta_1\right)\cos\theta_1  \sin 2\theta_2 \cos 2\theta_3-\left(1-10\cos^2\theta_1+\cos^2\theta_2+5\cos^2\theta_1\cos^2\theta_2
\right)\sin\theta_1\cos\theta_2\sin 2\theta_3}{12\sqrt{3}}\,,\nonumber \\
I_1  & = &  \frac{(-1)^{k_2-k_1}\sqrt{2}}{9} \cos \theta_1 \cos \theta_2 \left(\sin 2\theta_3(1+\sin^2\theta_1\cos^2\theta_2-2\cos^2\theta_1 \sin^2\theta_2)-\sin2\theta_1\sin\theta_2\cos2\theta_3 \right)\,,\nonumber \\
I_2  & = &  \frac{(-1)^{k_3-k_2}\sqrt{2}}{9}\left((\cos^2\theta_1+1)\sin2\theta_2\cos \theta_3-\sin 2\theta_1\cos \theta_2 \sin \theta_3\right) \left(\sin \theta_1 \cos \theta_3+\sin \theta_2 \sin \theta_3\cos \theta_1\right)\,.
\label{x1param1}
\end{eqnarray}

\begin{figure}[h!]
\begin{center}
\begin{tabular}{c}
\includegraphics[width=0.95\linewidth]{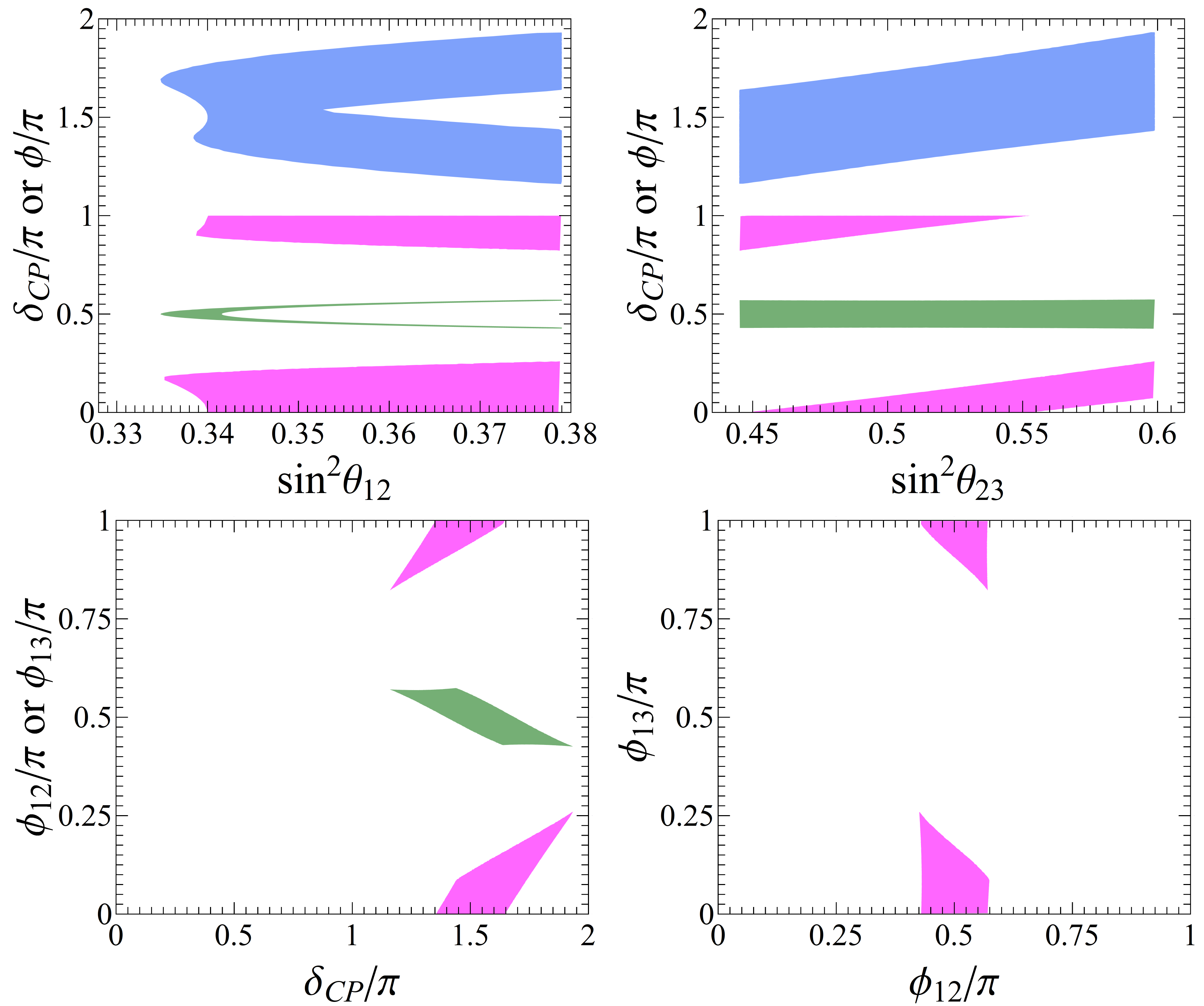}
\end{tabular}
\caption{ Correlations between mixing angles and CP phases from 
Eq.~\eqref{x1param1}. These hold when $X_1$ is preserved by the 
neutrino sector. In all panels the green and magenta regions correspond to the Majorana phases $\phi_{12}$ and $\phi_{13}$, respectively, while
the blue regions in the upper panels correspond to the Dirac phase $\delta_{CP}$.}
\label{fig:plots123}
\end{center}
\end{figure}

The resulting mixing parameter correlations are shown in Fig.~\ref{fig:plots123}

\subsection{Preserving  the $X_2$ CP symmetry}

Again using Eq.~\eqref{generalonecpneutrino} when $i=2$ we find that the lepton mixing matrix is given by
\begin{equation}
U_{lep} = U_{rTBM}\text{diag}(i, 1, i) O_{3\times3} Q_{\nu}\,.
\label{UX2}
\end{equation}

\begin{figure}[h!]
\begin{center}
\begin{tabular}{ccc}
\includegraphics[width=0.95\linewidth]{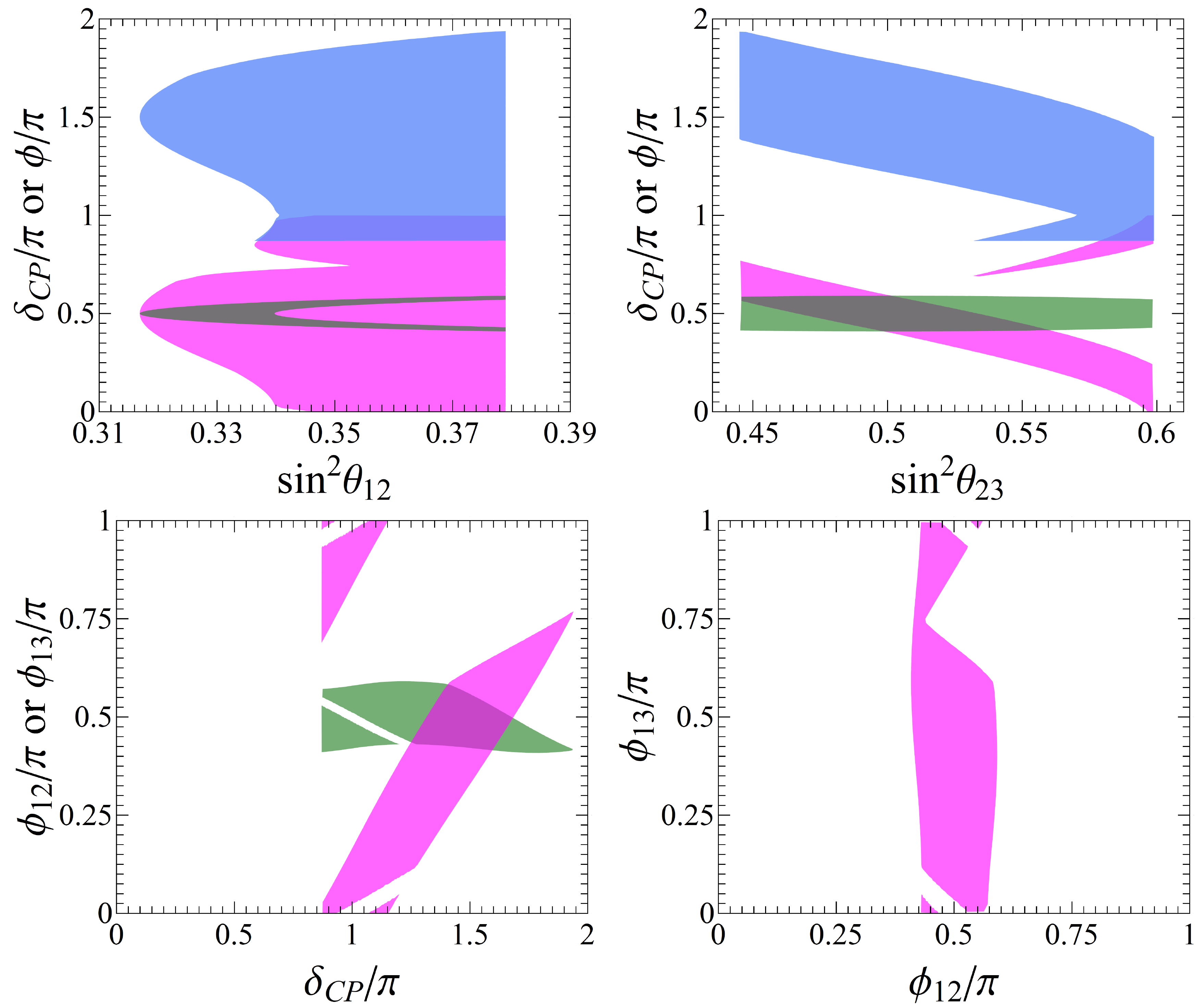}
\end{tabular}
\caption{ Correlations between mixing angles and CP phases from Eq.~\eqref{x2param1}. These hold when $X_2$ is preserved by the 
neutrino sector. In all panels the green and magenta regions correspond to the Majorana phases $\phi_{12}$ and $\phi_{13}$, respectively, while
the blue regions in the upper panels correspond to the Dirac phase $\delta_{CP}$.}
\label{plots223}
\end{center}
\end{figure}

As a consequence, the mixing parameters are given as
\begin{eqnarray}
\sin^2 \theta_{13}  & = & \frac{1}{3} \left(\cos^2 \theta_2 \sin^2\theta_1+2 \sin^2\theta_2\right)\,, \nonumber \\
\sin^2 \theta_{12}& = & \frac{(\cos\theta_1\cos\theta_3-\sin\theta_1\sin\theta_2\sin\theta_3)^2+2\cos^2\theta_2\sin^2\theta_3}{(1+\cos^2\theta_1)\cos^2\theta_2+1}\nonumber \\
\sin^2 \theta_{23}  & = &  \frac{1}{2}-\frac{2 \sqrt{3} \sin 2\theta_2 \cos \theta_1}{\cos 2\theta_1+2\left(\cos^2\theta_1+1\right) \cos 2\theta_2+7}\,,\nonumber \\
J_{CP}  & = &\frac{\left(10 \sin^2\theta_1\sin \theta_2 +\left(3 \cos 2\theta_1+5\right)\sin 3\theta_2 \right)\sin 2\theta_3 + 8 \sin 2\theta_1 \cos 2\theta_2 \cos 2\theta_3}{48 \sqrt{6}}\,, \\
I_1  & = &  \frac{(-1)^{k_2-k_1}\sqrt{2}}{9} \cos \theta_1 \cos \theta_2 \left(\sin 2\theta_3(2\cos^2\theta_1 \sin^2\theta_2-\sin^2\theta_1 \cos^2\theta_2-1)+\sin2\theta_1\sin\theta_2\cos2\theta_3 \right)\,,\nonumber \\
I_2  & = &  \frac{(-1)^{k_3-k_1}\sqrt{2}}{9}\left(\sin 2\theta_1\cos \theta_2 \sin \theta_3-(\cos^2\theta_1+1)\sin2\theta_2\cos \theta_3\right) \left(\sin \theta_1 \cos \theta_3+\sin \theta_2 \sin \theta_3\cos \theta_1\right)\,. \nonumber
\label{x2param1}
\end{eqnarray}
Various resulting mixing parameter correlations are shown in Fig.~\ref{plots223}.

\subsection{Preserving the $X_3$ CP symmetry}
%
For the case of $i=3$, we can see the $X_3$ is exactly the $\mu-\tau$ reflection symmetry. The lepton mixing matrix is of the form
\begin{eqnarray}
U_{lep} = U_{rTBM}\text{diag}(i, i, 1) O_{3\times3} Q_{\nu}\,,
\end{eqnarray}
where $U_{rTBM}\text{diag}(i, i, 1)$ can be decomposed as
\begin{eqnarray}
U_{rTBM}\text{diag}(i,i,1)=
\left(\begin{matrix}
-i&0&0\cr
0&-\frac{i}{\sqrt{2}}&\frac{1}{\sqrt{2}}\cr
0&\frac{i}{\sqrt{2}}&\frac{1}{\sqrt{2}}\cr
\end{matrix}\right)
\left(\begin{matrix}
-\sqrt{\frac{2}{3}}&-\frac{1}{\sqrt{3}}&0\cr
 \frac{1}{\sqrt{3}}&-\sqrt{\frac{2}{3}}&0\cr
 0&0&1\cr\end{matrix}\right)\,.
\end{eqnarray}
The constant matrix on the right side can be absorbed into the orthogonal matrix $O_{3\times3}$ by redefining the parameters $\theta_1$, $\theta_2$ and $\theta_3$. For simplicity we use the same notation for the reparameterized $O_{3\times3}$ so that
\begin{eqnarray}
U_{lep}=
\left(\begin{matrix}
-i&0&0\cr
0&-\frac{i}{\sqrt{2}}&\frac{1}{\sqrt{2}}\cr
0&\frac{i}{\sqrt{2}}&\frac{1}{\sqrt{2}}\cr
\end{matrix}\right)
O_{3\times3} Q_{\nu}\,.
\end{eqnarray}
Then we can read out the lepton mixing parameters as follow, 
\begin{eqnarray}
\nonumber&\sin^2 \theta_{13} = \sin^2\theta_2\,,\qquad
\sin^2 \theta_{12} = \sin^2\theta_3\,,\qquad
\sin^2 \theta_{23} = \frac{1}{2}\,,&\\
&\sin\delta_{CP}=\text{sign}(\sin\theta_2\sin2\theta_3)\,,\qquad
\phi_{12}=\frac{k_1-k_2}{2}\pi\,,\qquad
\phi_{13}=\frac{k_1-k_3}{2}\pi &\,.
\end{eqnarray}
Obviously, both $\theta_{23}$ and $\delta_{CP}$ are maximal, while the mixing angles $\theta_{12}$ and $\theta_{13}$ are unconstrained.  Both Majorana phases $\phi_{12}$ and $\phi_{13}$ also take conserved values.

\subsection{Preserving the $X_4$ CP symmetry}
%
Finally, notice that the $X_4$ symmetry is just the trivial symmetry of diagonal phases. 
Imposing only the $X_4$ symmetry indeed leads to leptonic mixing matrices consistent with all experimental observations.
However, in this case the neutrino mixing matrix will be a completely arbitrary orthogonal matrix ($\delta_{CP} = 0,\pi$ and no prediction for the mixing angles) while the Majorana phases will be simply $\pm \pi/2$, 0, $\pi$. This can also be seen from Eq.~\eqref{generalonecpneutrino} when $i=4$.

\section{Summary and discussion}

In this paper we have explored the CP symmetries admitted by the Tri-Bi-Maximal (TBM) mixing matrix. 
Using these CP symmetries as guidance, we have constructed several realistic variants of the TBM ansatz. Depending on the type and number of generalized CP symmetries imposed, we have obtained several realistic mixing matrices, all of which are related with the original TBM matrix. 
One of these variants is the recently discussed gTBM matrix in Ref.~\cite{Chen:2018eou}.
The correlations between solar and reactor angles are summarized in Fig.~\ref{fig:g1L}. 
The corresponding predictions for the atmospheric angle and the Dirac phase $\delta_{CP}$ are given in Fig.~\ref{fig:g1R}. 
These hold equally well irrespective of whether neutrinos are Majorana or Dirac-type.
Predictions for CP phases are collected in Figs.~\ref{fig:plots123} and \ref{plots223}. Their upper panels show predictions given in terms of the solar and atmospheric mixing angles, while the lower panels illustrate the results we obtain for the phase-phase correlations, both for Dirac as well as Majorana phases.
The predictions we have obtained can be tested in currently running as well as upcoming neutrino experiments.
Dedicated studies of the phenomenological implications of our predicted leptonic mixing matrix patterns will be taken up elsewhere.

\section{Acknowledgments}

This work is supported by National Natural Science Foundation of China under Grant Nos 11522546, 1183501 and 11847240 and China Postdoctoral Science Foundation under Grant Nos 2018M642700 and the Spanish grants FPA2017-85216-P (AEI/FEDER, UE), SEV-2014-0398 and PROMETEOII/2018/165 (Generalitat  Valenciana). S.C.C is also supported by the FPI grant BES-2016-076643.


\begin{appendix}


\section{Diagonalization of $3\times 3$ real symmetric matrix}
\label{sec:o33}
In this section, we would like to discuss how to diagonalize the matrix analogue to in Eq.~\eqref{eq:hatm13}. Consider diagonalize the following matrix:
\begin{equation}
\widetilde{M}=
\left(\begin{array}{ccc}
 m_1 & a & b \\
 a & m_2 & c \\
 b & c & m_3
\end{array}\right)\,.
\end{equation}
The eigenvalues of this matrix can be obtained from the formula of extracting roots on cubic equation with three different real roots.
The characteristic polynomial is
\begin{eqnarray}
\nonumber
&&\lambda^{3}-(m_{1}+m_{2}+m_{3})\lambda^{2}+(m_{1}m_{2}+m_{1}m_{3}+m_{2}m_{3}-a^{2}-b^{2}-c^{2})\lambda\\
&&~~+(-m_{1}m_{2}m_{3}+a^{2}m_{3}+b^{2}m_{2}+c^2m_{1}-2abc)=0\,.
\end{eqnarray}
For simplicity we define
\begin{eqnarray}
&&x\equiv -m_{1}-m_{2}-m_{3}\,, \nonumber \\
&&y\equiv m_{1}m_{2}+m_{1}m_{3}+m_{2}m_{3}-a^{2}-b^{2}-c^{2}\,, \nonumber \\
&&z\equiv -m_{1}m_{2}m_{3}+a^{2}m_{3}+b^{2}m_{2}+c^2m_{1}-2abc\,,
\end{eqnarray}
then we have
\begin{eqnarray}
&&\lambda_{1}=-\frac{x}{3}+2\sqrt{-\beta}\,S\,, \nonumber \\
&&\lambda_{2}=-\frac{x}{3}- \sqrt{-\beta}\,\left[S+\sqrt{3(1-S^{2})}\right]\,, \nonumber \\
&&\lambda_{3}=-\frac{x}{3}- \sqrt{-\beta}\,\left[S-\sqrt{3(1-S^{2})}\right]\,,
\end{eqnarray}
with
\begin{equation}
S=\cos\left[\frac{1}{3}\arccos \frac{\alpha}{(-\beta)^{3/2}}\right]\,,\quad
\alpha =-\frac{x^{3}}{27}-\frac{z}{2}+\frac{xy}{6}\,,\quad
\beta  = \frac{y}{3}-\frac{x^{2}}{9}\,.
\end{equation}
The orthogonal diagonal matrix of $M$ is given by
\begin{equation}
O_{3\times3}
=\left(\begin{array}{ccc}
\frac{(\lambda_{1}-m_{2})(\lambda_{1}-m_{3})-c^{2}}{C_{1}} &
\frac{(\lambda_{2}-m_{3})a+bc}{C_{2}}&
\frac{(\lambda_{3}-m_{2})b+ac}{C_{3}}\\
\frac{(\lambda_{1}-m_{3})a+bc}{C_{1}}&
\frac{(\lambda_{2}-m_{1})(\lambda_{2}-m_{3})-b^{2}}{C_{2}}&
\frac{(\lambda_{3}-m_{1})c+ab}{C_{3}}\\
\frac{(\lambda_{1}-m_{2})b+ac}{C_{1}} &
\frac{(\lambda_{2}-m_{1})c+ab}{C_{2}} &
\frac{(\lambda_{3}-m_{1})(\lambda_{3}-m_{2})-a^{2}}{C_{3}}
\end{array}\right)\,.
\end{equation}
with
\begin{eqnarray}
\nonumber
&&C_{1}=\sqrt{\big[(\lambda_{1}-m_{3})a+bc\big]^{2}+\big[(\lambda_{1}-m_{2})b+ac\big]^{2}+\big[(\lambda_{1}-m_{2})(\lambda_{1}-m_{3})-c^{2}\big]^{2}}\,,\\
\nonumber
&&C_{2}=\sqrt{\big[(\lambda_{2}-m_{3})a+bc\big]^{2}+\big[(\lambda_{2}-m_{1})c+ab\big]^{2}+\big[(\lambda_{2}-m_{1})(\lambda_{2}-m_{3})-b^{2}\big]^{2}}\,,\\
\nonumber
&&C_{3}=\sqrt{\big[(\lambda_{3}-m_{2})b+ac\big]^{2}+\big[(\lambda_{3}-m_{1})c+ab\big]^{2}+\big[(\lambda_{3}-m_{1})(\lambda_{3}-m_{2})-a^{2}\big]^{2}}\,.
\end{eqnarray}
Such that
\begin{equation}
O_{3\times3}^T\widetilde{M} O_{3\times3} = \left( \begin{array}{ccc}
\lambda_{1} & 0 & 0 \\
0 & \lambda_{2} & 0 \\
0 & 0 & \lambda_{3} \\
\end{array}\right)\,.
\end{equation}
\end{appendix}

\newpage

\bibliographystyle{utphys}
\bibliography{bibliography}
\end{document}